\documentclass[12pt,nofootinbib,preprint]{revtex4}
\pdfoutput=1

\usepackage{epsf, array, color}

\usepackage{amsmath}
\usepackage{amssymb}
\usepackage{anysize}
\usepackage{bold-extra}
\usepackage{xcolor}
\usepackage{enumerate}
\usepackage{fancyhdr}
\usepackage{graphicx}

\definecolor{darkblue}{rgb}{0.0, 0.0, 0.55}
\definecolor{forestgreen}{rgb}{0.13, 0.55, 0.13}
\usepackage[colorlinks=true,linktocpage=true,linkcolor=darkblue,citecolor=forestgreen,urlcolor=forestgreen]{hyperref}

\def\lsim{\mbox{\raisebox{-.6ex}{~$\stackrel{<}{\sim}$~}}}

\usepackage{amsfonts}
\usepackage{comment}
\usepackage{epigraph}
\usepackage[utf8]{inputenc}
\usepackage{slashed}
\usepackage{bbm}
\usepackage{cancel}
\usepackage{epsf, array, color}

\newcommand{\Op}{{\cal O}}
\newcommand{\nc}{\newcommand}
\nc{\vp}{H}
\nc{\tvp}{\widetilde{\phi}}
\nc{\vpj }{\mbox{${\vp^\dag i\,\raisebox{2mm}{\boldmath ${}^\leftrightarrow$}\hspace{-4mm} D_\mu\,\vp}$}}
\nc{\vpjt}{\mbox{${\vp^\dag i\,\raisebox{2mm}{\boldmath ${}^\leftrightarrow$}\hspace{-4mm} D_\mu^{\,a}\,\vp}$}}

\begin{document}

\preprint{YITP-SB-20-18}

\title{\boldmath Putting SMEFT Fits to Work}

\author{Sally Dawson$^{1}$}
\email{dawson@bnl.gov}

\author{Samuel Homiller$^{2}$}
\email{samuel.homiller@stonybrook.edu}

\author{Samuel D. Lane$^{1,3}$}
\email{samuel.lane@ku.edu}

\affiliation{
$^{1}$Department of Physics, Brookhaven National Laboratory, Upton, New York 11973~ U.S.A.\\
$^{2}$C. N. Yang Institute for Theoretical Physics, Stony Brook University, 
 Stony Brook, New York 11794~ U.S.A.\\
 $^{3}$Department of Physics and Astronomy, University of Kansas, Lawrence, Kansas, 66045~ U.S.A
 }

\begin{abstract}

The Standard Model Effective Field Theory (SMEFT) provides a consistent framework for comparing precision measurements at the LHC to the Standard Model. The observation of statistically significant non-zero SMEFT coefficients would correspond to physics beyond the Standard Model (BSM) of some sort.  A more difficult question to answer is what, if any, detailed information about the nature of the underlying high scale model can be obtained from these measurements.  In this work, we consider the patterns of SMEFT operators present in five example models and discuss the assumptions inherent in using global fits to make BSM conclusions.  We find that including renormalization group effects has a 
significant impact on the interpretation of the results. As a by-product of our study, we present an up-dated global fit to SMEFT coefficients  in the Warsaw basis including some next-to-leading order QCD corrections in the SMEFT theory.

\end{abstract}

\maketitle

\tableofcontents

\clearpage
\section{Introduction}
\label{sec:intro}

With the Higgs discovery in hand and the Standard Model (SM) field content complete, one of the primary goals of the LHC is
to make precise measurements of SM processes, with the hope of testing its limitations. 
As the search for new particles has been unsuccessful as yet, much attention has shifted towards precision measurements of Higgs processes.
In this direction, the SMEFT (SM effective field theory) framework becomes very useful. In the SMEFT, deviations from the SM are parameterized by a tower of $SU(3)\times SU(2)_L\times U(1)_Y$ invariant higher dimension operators containing only SM fields.
The SMEFT is useful because it provides a consistent, gauge-invariant theoretical interpretation of the data, in which higher order corrections can be included, and connects Higgs data with electroweak precision observables at the $Z$ and $W$ poles, diboson data, and top quark measurements.
There have been numerous global fits to LHC and LEP data, yielding limits on the allowed values of the SMEFT coefficients~\cite{Biekotter:2018rhp, Grojean:2018dqj, Ellis:2018gqa, Almeida:2018cld, DiVita:2017eyz}, but thus far all of these fits are totally consistent with the Standard Model, further demonstrating the robustness of the SM.

Ultimately, however, one hopes that the SMEFT is parameterizing some high scale   physics beyond the Standard Model (BSM), and it is of interest to understand how these global fits should be interpreted in this context.
The goal of this work is to consider just such an interpretation in detail for a set of simple benchmark models. 
We consider a sample set of UV complete models where the BSM physics contains particles at mass scales above the weak scale, $\Lambda \gg M_Z$.
Each model makes a prediction for the SMEFT coefficients at the high scale, $C_i(\Lambda)$, and typically only a small subset of dimension-6 operators are generated~\cite{deBlas:2017xtg, Henning:2014wua, Bakshi:2018ics}. 
As we will see, all of our models predict particular relationships between the different coefficients generated, and we will explore the differences in fitting with these particular patterns as opposed to general values of the coefficients.
We will mostly restrict ourselves to tree-level matching between the BSM physics and the SMEFT at the scale $\Lambda$, but we will also consider the renormalization group running to evolve the coefficients at leading logarithm to the weak scale~\cite{Grojean:2013kd, Chen:2013kfa}, where the predictions can be compared with fits to the data.
Since many of our benchmark models generate operators that are well constrained by electroweak precision observables, including these leading logarithmic effects will significantly change the interpretations of the fits.
These considerations are an important first step in understanding what we can learn about possible UV complete models, and, if a deviation from the Standard Model is observed, how we can discriminate between them. This goal is sometimes referred to as the ``Higgs Inverse Problem''.

Understanding the sensitivity to BSM physics through the extraction of SMEFT coefficients is also useful for comparing the reach of future accelerators~\cite{deBlas:2019rxi} and our calculations are  part of an extended effort to understand the complementarity of the direct observation of new particles with precision measurements~\cite{Gupta:2012mi}. 
In addition to Higgs signal strength data, we include theory predictions with NLO QCD corrections for $VV$ and $VH$ ($V = W, Z$) 
production~\cite{Baglio:2020oqu}, and the leading logarithmic NLO QCD and electroweak  corrections to $Z$ and $W$ pole observables computed in the full SMEFT~\cite{Dawson:2019clf}.
As a by-product of our study, we obtain an update of the global fit to SMEFT coefficients in the Warsaw basis.

We begin by describing how we match our benchmark models to the SMEFT, including the effects of
renormalization group evolution  (RGE) of the Wilson coefficients down to the weak scale. We then summarize each of the models in turn in Section~\ref{sec:modelsConsidered}. The operators generated in each model are summarized in Section~\ref{sec:summary}.
We then perform a series of fits customized for each model in Section~\ref{sec:results}, and discuss how the SMEFT fit results can differ depending on the correlations present in the underlying high scale model. 
Here we will further emphasize the role of the renormalization group evolution of the coefficients in the interpretation of SMEFT fit results. 
Finally, we conclude with our updated global fit  in Section~\ref{sec:global} and a discussion of future directions in this type of study in Section~\ref{sec:conc}.

\section{Matching Models to the SMEFT}
\label{sec:models}

In this section, we lay out our benchmark models, and tabulate the relevant SMEFT coefficients obtained in the decoupling limit of these models. The models are chosen to be simple but representative BSM models with new particles at the $\sim$ few TeV scale with only a small set of unknown parameters.
The full set of dimension-6 operators that we will  consider, ignoring flavor,  is given in Table~\ref{tab:opdef}.

\begin{table}[t] 
\centering
\renewcommand{\arraystretch}{1.5}
\begin{tabular}{|c|c||c|c||c|c|} 
\hline
${\Op_{ll}}$                   & $(\bar l_L \gamma_\mu l_L)(\bar l_L \gamma^\mu l)_L$  &    
  ${\Op}_{H W B}$ 
 &$ (\vp^\dag \tau^a \vp)\, W^a_{\mu\nu} B^{\mu\nu}$  &
$\Op_{\vp D}$   & $\left(\vp^\dag D^\mu\vp\right)^* \left(\vp^\dag D_\mu\vp\right)$ 
\\
\hline 
   ${\Op}_{ He}$  &   $(\vpj) (\overline {e}_R\gamma^\mu e_R)$  & 
   ${\Op}_{H u}$ & $(\vpj) (\overline {u}_R\gamma^\mu u_R)$ &
  ${\Op}_{H d}$
       & $(\vpj) (\overline {d}_R\gamma^\mu d_R)$
  \\ \hline 
             ${\Op}_{ Hq}^{(3)}$ & $(\vpjt)(\bar q_L \tau^a \gamma^\mu q_L)$  &${\Op}_{H q}^{(1)}$
      &$(\vpj)(\bar q_L \gamma^\mu q_L)$  &
 ${\Op_{\vp l}^{(3)}}$      & $(\vpjt)(\bar l_L \tau^a \gamma^\mu l_L)$
 \\
\hline
${\Op}_{H l}^{(1)}$ &  
$(\vpj)(\bar l_L  \gamma^\mu l_L)$ & 
${\Op}_{H\square}$ & $(H^\dagger H)\square (H^\dagger H)$  &
${\Op}_{eH}$ & $(H^\dagger H) \bar l_L {\tilde H} e_R$
\\
\hline
${\Op}_{HG}$ & $(H^\dagger H) G_{\mu\nu}^A G^{\mu\nu,A}$  &
${\Op}_{uH}$ &   $(H^\dagger H)({\overline q}_L{\tilde {H}}u_R)$
 &
${\Op}_{dH}$ & $(H^\dagger H)({\overline q}_L{ {H}}d_R)$
\\
\hline 
${\Op}_{HB}$ & $(H^\dagger H) B_{\mu\nu} B^{\mu\nu}$ &
${\Op}_{HW}$  & $(H^\dagger H) W_{\mu\nu}^a W^{\mu\nu,a}$  &
${\Op_W}$  & $\epsilon_{abc}W_\mu^{\nu,a}W_\nu^{\rho,b}W_\rho^{\mu,c}$
\\
\hline 
${\Op}_H$ & $(H^\dagger H)^3$ &
& &
 &
\\
\hline
\end{tabular}'
\caption{Dimension-6 operators in the Warsaw basis that are considered in this study.  The fermion labels $u, d, e$ refer generically  to all $3$ generations and similarly for the quark, $q$, and lepton, $l$, doublets.\label{tab:opdef}}
\end{table}

\subsection{Matching Procedure and Renormalization Group}
\label{sec:matching}

Here we lay out our benchmark models, and tabulate the relevant SMEFT coefficients obtained in
the decoupling limits of the models.  The models are chosen to be well studied  BSM models with new particles at
the UV scale and are quite simple models, involving a small set of unknown parameters.  
We assume that the new particles are at a high mass scale and integrate them out of the theory using the equations of motion at tree level, although for future studies it would be of interest to perform the matching at one-loop, since  one-loop matching typically generates a much
richer spectrum of SMEFT operators than does the tree level matching~\cite{Gorbahn:2015gxa, Henning:2014wua, Jiang:2018pbd}. 
This  procedure generates predictions for SMEFT coefficients at the mass scale $\Lambda$ corresponding to the new physics,
\begin{equation}
\mathcal{L} = \mathcal{L}_{SM}+\sum_i {C_i\over \Lambda^2}\Op_i^6+ \dots\, ,
\end{equation}
where we include the dimension-6 operators in the Warsaw basis~\cite{Dedes:2017zog, Grzadkowski:2010es}.   The operators
consist of all  of the $SU(3)\times SU(2)_L\times U(1)_Y$ operators that can be constructed from SM fields. 
Since we assume $\Lambda \gg M_Z$, we only consider the dimension-6 operators, $O_i^6$.
Some of the
models we consider generate effects only for third generation quarks, so we do not always assume flavor universality in the
quark sector.
The importance of the assumptions about flavor in the results of the global fits has been emphasized in Refs.~\cite{Falkowski:2019hvp, Grojean:2018dqj}.

We fit data from Higgs processes, diboson $WW$ and $WZ$  production, and electroweak precision observables (EWPOs), including
the $W$ mass and width, to the
patterns of SMEFT coefficients generated in our examples. 
For completeness, we define the operators appearing in this work in Table~\ref{tab:opdef} and we neglect flavor indices in this table, although we will include them in some of the models.  We define $H$ to be the $SU(2)_L$ doublet Higgs field with neutral component ${h+v\over\sqrt{2}}$ and,  in terms of the first generation,
$q_L^T=(u_L, d_L)$, $l_L^T=(\nu_L, e_L)$.   Our notation follows that of Ref.~\cite{Dedes:2017zog}.

At tree level, the $Z$ and $W$ pole observables depend on,
\begin{equation}
{\Op_{ll}},\quad
  {\Op}_{H W B}, \quad
\Op_{\vp D},\quad
 {\Op}_{H e},\quad
  {\Op}_{H d}, \quad
  {\Op}_{H u},\
  {\Op}_{H q}^{(3)}, \quad
  {\Op}_{H  q}^{(1)}, \quad
 {\Op_{\vp l}^{(3)}}, \quad
 {\Op}_{H  l}^{(1)},
\end{equation} 
and the EWPOs are sensitive to eight combinations of these operators~\cite{Dawson:2019clf, Corbett:2017qgl, Falkowski:2014tna, Berthier:2016tkq}, (at NLO they are sensitive to a combination
of $10$ operators).  We  also include the $2-$ loop contribution to $M_W$ generated by $\Op_H$~\cite{Degrassi:2014sxa, Kribs:2017znd}.

We consider tree level contributions from the  following operators to Higgs data,
\begin{eqnarray}
\Op_{HW},\quad 
\Op_{HB},\quad 
\Op_{HWB},\quad 
\Op_{HD},\quad 
\Op_{HG},\quad 
\Op_{H\square},\quad 
\Op_{Hl}^{(3)},\quad 
\Op_{He},\quad  \nonumber \\
\Op_{Hu},\quad 
\Op_{Hd},\quad 
\Op_{ll},\quad 
\Op_{Hq}^{(1)}, \quad
\Op_{Hq}^{(3)},\quad 
\Op_{eH},\quad
\Op_{uH},\quad
\Op_{dH}\, .
\end{eqnarray}
We also include the loop contributions to Higgs production and decay from $\Op_H$~\cite{Degrassi:2016wml, Gorbahn:2016uoy, Bizon:2016wgr}. 

Finally, the diboson $WW$ and $WZ$ data depend on $7$ effective couplings\footnote{See~\cite{Baglio:2018bkm} for
a convenient mapping from the effective interactions to the Warsaw basis.},
which involve the operators 
\begin{equation}
\Op_{HWB},\quad 
\Op_{HD},\quad 
\Op_W,\quad 
\Op_{Hl}^{(3)},\quad 
\Op_{ll},\quad 
\Op_{Hq}^{(1)},\quad 
\Op_{Hq}^{(3)},\quad 
{\Op}_{Hu},\quad 
\Op_{Hd} ,
\end{equation}
for a total of 19 operators involved in our study when we neglect flavor. For the vector-like quark models we consider,
only contributions to operators involving third generation quark interactions  are generated.  Most of the operators that contribute to the global fits  that we list above do not arise in
 the models we consider and we comment on the importance of this in  the concluding discussion.

The models we consider are chosen to illustrate various types of new physics and to demonstrate uncertainties and assumptions that are made when forming physics conclusions from the global fits. 
The models fall in two categories. 
\begin{itemize}
\item{Models with high scale scalar resonances:
We consider a real scalar  singlet model, both with and without a $Z_2$ symmetry, and a $2$ Higgs doublet model (2HDM)  in the decoupling limit.}

\item{Models with new particles in loops:
We consider two  models with vector-like quarks (VLQs): One with a color triplet fermion with charge $Q={2\over 3}$, and one with a color triplet,
$SU(2)_L$ doublet of quarks with
charge $Q=({2\over 3}, -{1\over 3})$.
We also briefly compare the results of the models with  vector-like quarks with  a model containing
a heavy color triplet scalar.}

\end{itemize}
There have been extensive studies in the literature computing SMEFT coefficients in these models. 
We summarize the models we consider below and the reader is referred to the original literature for further details.

The fits are performed in two different manners.  In the first approach, we match the coefficients at the UV scale $\Lambda$  to the model predictions and perform the global fit.
These fits are only sensitive to the ratios ${C_i\over \Lambda^2}$ and give no independent information about the UV scale.
We always make the identification that $\Lambda$ is the mass of the heavy particle that has been integrated out. 
In the second set of fits, we match the coefficients to the model predictions at $\Lambda$ and then use the renormalization group to evolve the coefficients to $M_Z$ before performing the fits.
The coefficients at the weak scale are then,
\begin{equation}
C_i(M_Z)=C_i(\Lambda) -{1\over 16\pi^2}{\dot C}_i\log\biggl({\Lambda\over M_Z}\biggr)\, , ~{\dot C}_i\equiv \gamma_{ij}C_j\, .
\end{equation}
A complete set of the relevant anomalous dimensions  in the Warsaw basis is in Refs.~\cite{Jenkins:2013zja,Jenkins:2013wua,Alonso:2013hga}.
In many cases, the RGE has a dramatic effect on the interpretation of the fits.
The EWPO and diboson data place strong constraints on $\Op_{HD}$, $\Op_{H\square}$,  and $\Op_{HWB}$.
The contributions involving the first and second generation quarks to $\Op_{Hq}^{(3)}$, and $\Op_{Hq}^{(1)}$
are strongly constrained by Higgstrahlung data, while the $Z\rightarrow b {\overline b}$ data constrain these operators with $3^{rd}$ generation quarks.  In models where these coefficients are generated by RGE (even when they are not present at the matching scale), the constraints and the interpretations change dramatically~\cite{Chen:2013kfa}.    
 
To illustrate the importance of including the RGE when fitting to UV complete models, we consider the 
strongly constrained operators,  $\Op_{HD}$, $\Op_{H\square}$,   $\Op_{Hq}^{(1)}$ and $\Op_{Hq}^{(3)}$ ~\cite{Baglio:2020oqu}.
$\Op_{HD}$ is not generated at tree level in any of the models we consider, while  $\Op_{H\square}$ arises at tree level in the singlet model.
Including only contributions from terms that are generated at tree level in the models we consider (see Table~\ref{tab:sumlag}),  
\begin{eqnarray}
{\dot C}_{HD} &=& 
{8\over 3}g^{\prime~2}
\biggl[2C_{Ht}-C_{Hb} +\big(C_{Hq}^{(1)}\big)_{33} \biggr] +{20\over 3}g^{\prime ~2}C_{H\square} \nonumber \\
&& -24\biggl[Y_t^2C_{Ht}-Y_b^2C_{Hb}+Y_bY_tC_{Htb}\biggr]\nonumber \\
&&+24\biggl(Y_t^2-Y_b^2\biggr)\big(C_{Hq}^{(1)}\big)_{33}  \nonumber \\
{\dot C_{H\square}} 
&=& 
6 g^2 \big(C_{Hq}^{(3)}\big)_{33} +{2\over 3}g^{\prime~2}\biggl[2C_{Ht}-C_{Hb}+\big(C_{Hq}^{(1)}\big)_{33} \biggr]
\nonumber \\ 
&&+\biggl[
-{4\over 3}g^{\prime~2}-4g^2+12\biggl(Y_t^2+Y_b^2\biggr)+4Y_\tau^2\biggr] C_{H\square}
\nonumber \\ && 
- 6\biggl[(Y_b^2-Y_t^2)\big(C_{Hq}^{(1)}\big)_{33} 
 +3(Y_b^2+Y_t^2)\big(C_{Hq}^{(3)}\big)_{33}+Y_t^2 C_{Ht}-Y_b^2 C_{Hb}-2Y_bY_tC_{Htb}\biggr]\nonumber \\
\nonumber \\
\big({\dot C}_{Hq}^{(3)}\big)_{33}&=&
3 \biggl[ Y_b^2 - Y_t^2 \biggr] \big(C_{Hq}^{(1)}\big)_{33}
+ \biggl[ -{11\over 3} g^2 + 8 Y_t^2 + 8 Y_b^2 + 2 Y_\tau^2  \biggr] \big(C_{Hq}^{(3)}\big)_{33} \nonumber \\ &&
-{1\over 6}\biggl[3 Y_t^2 + 3Y_b^2 -g^2 \biggr] C_{H\square} 
\nonumber \\ 
\big({\dot C}_{Hq}^{(1)}\big)_{33}
&=&
 \biggl[ {5 \over 9}g^{\prime ~2}  + 10 Y_t^2 + 10 Y_b^2 + 2 Y_\tau^2  \biggr] \big(C_{Hq}^{(1)}\big)_{33} 
- 9 \bigg[Y_t^2-Y_b^2\bigg] \big(C_{Hq}^{(3)}\big)_{33} 
\nonumber \\ &&
-{1\over 2}\biggl[{ g^{\prime ~2} \over 9}+ Y_b^2 - Y_t^2 \biggr] C_{H\square} 
-(Y_t^2 + {4\over9}g^{\prime ~2} ) C_{Ht} 
-(Y_b^2 + {2\over9}g^{\prime ~2} ) C_{Hb} \, ,
\label{eq:regsum}
\end{eqnarray}
where $C_{Ht}, C_{Hb}, C_{Htb}, C_{H\tau},   (C_{Hq}^{(1)})_{33}$ and $(C_{Hq}^{(3)})_{33}$ correspond to operators including only the third generation fermions. 
See~\cite{Jenkins:2013zja,Jenkins:2013wua,Alonso:2013hga} for a full expression.
We note that $\Op_{HD}$ arises from the RGE in the $(TB)$ VLQ model and in  the singlet model, and so we expect  quite different results in these models when  the RGE is included.  Similarly, 
$(\Op_{Hq}^{(1)})_{33}$ and $(\Op_{HQ}^{(3)})_{33} $ occur at tree level in the $T$ VLQ model, and RGE generates $\Op_{H\square}$ at the weak scale which is strongly constrained.  On the other hand, the operators generated at tree level  in the 2HDM do not contribute to the RG evolution of  $C_{HD}$ or $C_{H\square}$ and we will see that RGE has a relatively minor effect on the interpretation of this model.  In our numerical
results, we include the complete RGE of all the operators that contribute to our fits.

\subsection{Models}
\label{sec:modelsConsidered}

\paragraph{Singlet Scalars\\}
One of the simplest extensions to the Standard Model is obtained by adding an additional scalar that is a  singlet under the SM gauge group. 
The scalar  potential can be constructed  both with and without a $Z_2$ symmetry. 
The case without a $Z_2$ symmetry  is particularly interesting because it can accommodate a first order electroweak phase transition for some values of the parameters~\cite{Profumo:2014opa,Chen:2017qcz}.  
Using the  classical equations of motion~\cite{Henning:2014wua,deBlas:2014mba,Gorbahn:2015gxa,Dawson:2017vgm}, 
the heavy scalar can be integrated out at tree level, generating  the SMEFT operators,
$\Op_H$ and $\Op_{H\square}$ with coefficients, 
\begin{eqnarray}
 {v^2\over \Lambda^2} C_{H\square}&=& -{1\over 2}\tan^2\theta\, ,\nonumber \\
 C_H &=& -C_{H\square}\biggl(\tan\theta{m\over 3 v}-\kappa\biggr)\, ,
\end{eqnarray}
where $\theta$ is the mixing angle between the SM-like Higgs boson, $h$, and the new heavy scalar and $\kappa$ and $m$ are Lagrangian parameters that are limited by the requirement that the electroweak minimum be the lowest minimum of the potential~\cite{Chen:2014ask}.
  In this model, the SM-like Higgs couplings to SM particles are uniformly suppressed by  a factor of $\cos\theta$ and 
for the case with a $Z_2$ symmetry, there is a cancellation implying $C_H=0$~\cite{Dawson:2017vgm,Gorbahn:2015gxa}.
Details of the model are in Appendix~\ref{sec:scalars}.\\

\paragraph{A Second Higgs Doublet\\}
The 2 Higgs doublet model (2HDM)  has been extensively studied in the literature, and in the limit that the new Higgs bosons are
much heavier than the SM-like Higgs bosons, the Higgs couplings approach those of the SM~\cite{Gunion:2002zf}.   This is the alignment
limit, $\cos(\beta-\alpha)\rightarrow 0$.  In the exact alignment limit, SMEFT operators are not generated at tree level, but 
first appear at 1-loop.  Away from the alignment limit, $\cos(\beta-\alpha) \ll 0$, tree level contributions to the Higgs-Yukawa
couplings are generated, along with a correction to the Higgs tri-linear coupling.  
To linear
order in $\cos(\beta-\alpha)$, the SMEFT coefficients that affect  Higgs couplings 
to fermions, $f$, are~\cite{Gorbahn:2015gxa,Brehmer:2015rna,Belusca-Maito:2016dqe,deBlas:2017xtg}, 
\begin{equation}
 {v^2 C_{fH}\over \Lambda^2} =-Y_f\eta_f{\cos(\beta-\alpha)\over\tan\beta} \, ,
\end{equation}
where $Y_f=\sqrt{2}m_f/v$ and $\eta_f$ distinguishes the type of 2HDM and is defined in Appendix~\ref{sec:2hdm}.
The corrections to the $WWh$ vertex  are ${\cal{O}}[\cos^2(\beta-\alpha)]$ and are neglected in
our approximation, since they are formally of dimension-8.
The correction to the $\Op_H$ operator scales slightly differently,
\begin{equation}
{v^2 C_H\over \Lambda^2}={\cos^2(\beta-\alpha) M^2\over v^2}\, ,
\end{equation}
where $M$ is the common mass of the heavy decoupled scalars near the alignment limit. 
We stress that  our results are only valid near  the alignment limit, where $\cos(\beta-\alpha) \ll 1$and terms of
${\cal  O}[\cos^2(\beta-\alpha)]$ are systematically neglected.
Further details of the model are found in Appendix~\ref{sec:2hdm}.\\

\paragraph{Colored Extensions of the SM: an $SU(3)$ Triplet, $SU(2)_L$ Singlet Fermion\\}
We consider a charge $Q={2\over 3}$ color triplet, $SU(2)_L$ singlet fermion, $T$, and call this the $T$ VLQ  (vector-like quark) model
and  assume that this new quark only couples to the top quark, but not to the lighter quarks.  The model is parameterized by
$3$ parameters:  $m_t$ and $M_T$ are the masses of the  physical top and  new heavy charge  $Q={2\over 3}$  fermion respectively, and $s_L^t$ is the sine of a mixing angle
that defines the mixing between the  left-handed charge ${2\over 3}$ quarks.  
Integrating out the heavy fermion generates the SMEFT coefficients involving the third generation quarks only~\cite{Chen:2017hak,Dawson:2012di,delAguila:2000aa,deBlas:2017xtg,Chen:2014xwa}
\begin{eqnarray}
{v^2\over \Lambda^2} \big(C_{Hq}^{(1)}\big)_{33}&=& - {v^2\over \Lambda^2}\big(C_{Hq}^{(3)}\big)_{33}= 
\biggl({1\over 2 Y_t} \biggr){v^2\over \Lambda^2}C_{tH} = {(s_L^t)^2\over 2}
\label{eq:topsing}
\end{eqnarray}
where 
\begin{eqnarray}
\big(\Op_{Hq}^{(1)}\big)_{33}&=& (\vpj)(\bar q_L^3  \gamma^\mu q_L^3)  \nonumber \\
\big(\Op_{Hq}^{(3)}\big)_{33}&=& (\vpjt)(\bar q_L^3 \tau^a \gamma^\mu q_L^3)  \nonumber\\
\Op_{tH}&=& (H^\dagger H)({\overline q}_L^3{\tilde {H}}t_R)\, ,
\end{eqnarray}
 ${\overline q}_L^3=({\overline t}_L, {\overline b}_L)$, and the scale $\Lambda$ is identified with $M_T$.  
The corresponding coefficients for the first 2 generations are zero in this model.

Although we perform the matching at tree level, we also include $\Op_{HG}$ since it could potentially make a significant
contribution to Higgs production through gluon fusion,
\begin{equation}\label{eq:vlqt_chg}
{v^2\over \Lambda^2} C_{HG}={\alpha_s\over 8\pi}(s_L^t)^2\biggl(F_{1/2}(M_T)-F_{1/2}(m_t)\biggr)\sim {\cal{O}}\biggl({\alpha_s (s_L^t)^2m_t^2\over M_T^2}\biggr)\,,
\end{equation}
where $F_{1/2}$ is defined in Appendix~\ref{sec:vlqsing}. 
For $M_T=1\,\textrm{TeV} ~(2\,\textrm{TeV}) $,  $F_{1/2}(M_T)-F_{1/2}(m_t)\sim- .010~ (-.011)$ and the top and $T$ contributions approximately cancel in
the large $m_t$ limit. A summary of the model is in Appendix~\ref{sec:vlqsing}.\\

\paragraph{Colored Extensions of the SM: an $SU(3)$ Triplet, $SU(2)_L$ Doublet Fermion\\}
We next consider a model with an $SU(2)_L$ doublet, color triplet pair of vector-like fermions.
We term this the $(TB) $ VLQ model.  
At tree level, the $(TB)$ doublet generates $O_{Ht},~O_{Hb},~O_{Htb},~O_{tH},$ and $O_{bH}$, where the operators are defined in Table~\ref{tab:opdef} in terms of $3^{rd}$ generation fermions. In the decoupling limit, $M_T, M_B \gg M_Z$, 
\begin{eqnarray}
 {v^2\over \Lambda^2}C_{Ht}&=&- {v^2\over \Lambda^2}{C_{tH}\over Y_t}=-(s_R^t)^2\nonumber \\
 {v^2\over \Lambda^2} C_{Hb}&=&{v^2\over \Lambda^2}{C_{bH}\over Y_b}=(s_R^b)^2 \nonumber \\
  {v^2\over \Lambda^2} C_{Htb}&=& 2s_R^t s_R^b,
\end{eqnarray}
where $s_R^t$ and $s_R^b$ define the   mixing between  the top and bottom quarks with T and B respectively in the right-handed sector.
At one-loop, $\Op_{HG}$ is also generated, 
\begin{eqnarray}\label{eq:vlqtb_chg}
 {v^2\over \Lambda^2}  C_{HG}&=&{\alpha_s\over 8 \pi} \biggl((s_R^t)^2\biggl[F_{1/2}(M_T)-F_{1/2}(m_t)\biggr]+(s_R^b)^2F_{1/2}(M_B)\biggr)
 \nonumber \\
 &\sim &{\alpha_s\over 8 \pi}(s_R^b)^2 (.32)\, ,
\end{eqnarray}
where $M_T=M_B=1\,\textrm{TeV}$ in the last equation.    The approximate cancellation between the $t$ and $T$
contributions found in the $T$ VLQ model remains, but there is an additional contribution from $M_B$.
Details can be found in Appendix~\ref{sec:vlqdoub}.\\

\paragraph{Colored Extensions of the SM: an $SU(3)$ Triplet Scalar\\}
Finally, we consider a model with a color triplet complex scalar, $s$,  with charge $Q={2\over 3}$ and mass, $m_s$. At tree level, four fermion operators that
do not contribute to our global fit are generated, but no dimension-6 EFT operators arise.
At one loop, the colored scalar generates $\Op_{HG}$,
\begin{equation}
{v^2C_{HG}\over \Lambda^2}= -{\alpha_s\kappa v^2 \over 96 \pi m_s^2}\, ,
\end{equation}
where  $\kappa$ is the portal coupling, 
$(s\,s^*)\big( H^{\dagger}H\big)$,
and is defined in Appendix~\ref{sec:colorscalar}.
This is an example of a model where the only effect on single Higgs production is to rescale the rate and the SMEFT formalism is not necessary. 
The indirect consequences  of the colored scalar and the corresponding SMEFT effects can be searched for in Higgs plus jet or double Higgs production~\cite{Dawson:2015oha,Dawson:2015gka}.

\newpage
\subsection{Summary of Models}
\label{sec:summary}

In Table~\ref{tab:sumlag}, we summarize the coefficients generated by our benchmark models as described in the previous sections, expressing the results in terms of the physical parameters of these models when possible. 
The scale $\Lambda$ is consistently identified with the mass of the heavy particle in the model.
More precise definitions of these parameters are given in the Appendices.

\begin{table}[h]
\renewcommand{\arraystretch}{1.1}
\begin{tabular}{| c | c | c | c | c | c | c | } 
\hline
& ~${\text{Singlet}}_{\slashed{\mathbb{Z}}_2}$~
&~${\text{Singlet}}_{\mathbb{Z}_2}$~ 
&~2HDM~
&~T VLQ~
&~(TB) VLQ~ 
&~~~~~$s$~~~~~
\\
\hline\hline 
 ${v^2 C_H  \over \Lambda^2}$ & 
${\tan^2\theta\over 2}(\tan\theta{m\over 3 v}-\kappa)$ &
&
${\cos^2(\beta-\alpha) M^2\over v^2}$ &
&
&
\\
\hline
  ${v^2 C_{H\Box} \over \Lambda^2}$ & 
  $- {\tan^2\theta\over 2}$   & 
  $- {\tan^2\theta\over 2}$ &
  &
  &
  &
  \\
  \hline 
 ${v^2 C_{bH}\over \Lambda^2}$ & 
& 
&
$-Y_b\eta_b{\cos(\beta-\alpha)\over \tan\beta} $&
&
$ Y_b (s_R^b)^2$ &
\\
\hline
${v^2 C_{tH}\over \Lambda^2}$ 
&
&
& - $Y_t\eta_t{\cos(\beta-\alpha)\over \tan\beta} $
& ${Y_t (s_L^t)^2 }$
& $Y_t (s_R^t)^2$
&
\\
\hline
${v^2 C_{\tau H}\over \Lambda^2}$ & 
& 
&
$-Y_\tau\eta_\tau{\cos(\beta-\alpha)\over \tan\beta} $&
&
 &
\\
\hline
 ${v^2 (C_{Hq}^{(1)})_{33}\over \Lambda^2}$ 
&
&
&
&
${(s_L^t)^2 \over 2 } $
&
&
\\
\hline
 ${v^2 (C_{Hq}^{(3)})_{33}\over \Lambda^2}$ 
 &
 & 
 &
 & $- {(s_L^t)^2\over 2} $
 &
 &
 \\
 \hline
 ${v^2 C_{Hb}\over \Lambda^2}$ 
 &
 &
& 
&
& ${ ( s_R^b)^2 } $
 & 
 \\
 \hline 
${v^2 C_{Ht}\over \Lambda^2}$ 
&
&
&
&
&
$- {(s_R^t)^2 } $
&
\\
\hline
 ${v^2 C_{Htb}\over \Lambda^2}$
&
&
&
&
&
$2{s_R^t s_R^b}  $
&
\\
\hline
${v^2 C_{HG}\over \Lambda^2}$ 
& 
& 
&
& 
$-{\alpha_s (s_L^t)^2\over 8 \pi }(.02)$
&${\alpha_s (s_R^b)^2\over 8 \pi }(.65)$
& $ -{\alpha_s \kappa v^2 \over 96\pi m_s^2}$ \\
\hline 
\end{tabular}
\caption{Tree level SMEFT coefficients. 
We also list $C_{HG}$ which is generated at 1-loop in some models and give numerical values for $C_{HG}$ for heavy masses of $1\,\textrm{TeV}$.
In all cases, we assume the decoupling limit of the models and the parameters are defined in the appendices. 
Empty spaces correspond to operators not generated at tree level.
\label{tab:sumlag} }
\end{table}

\newpage
\section{Results}
\label{sec:results}

\subsection{Methodology}
\label{sec:direct}

We perform a series of fits to Higgs, diboson, and EWPO data with prior assumptions about the relationships
between SMEFT coefficients that are motivated by our example models.
We take as non-zero only those coefficients generated in a particular model and examine how that choice changes the fits and the interpretations of the fit results. 
The underlying goal is to see how the fits can potentially constrain the high scale models. 
All of our fits are performed to linear order in the SMEFT coefficients.
 
The EWPO fits use the data given in Table III of~\cite{Dawson:2019clf}. 
This table uses the most accurate SM theoretical predictions available, typically at NNLO accuracy,  along with the tree level SMEFT contributions.
The LEP-2 data from $e^+e^-\rightarrow W^+W^-$ as parameterized in Ref.~\cite{Ellis:2018gqa} is included in the global fits. 
Using the results in the appendix of Ref.~\cite{Dawson:2019clf}, it is straightforward to generalize the leading order SMEFT contributions to the $\chi^2$ given there to allow for the quark operators to have different (but diagonal) couplings to the generations.
In particular the coefficients involving the $b$ quark can be different from those of the lighter fermions, which is relevant for the $T$ and ($TB$) VLQ models. 
We assume that the lepton generations all couple identically.
The contribution of the operator $\Op_H$ occurs at 2-loops in the SM and is included using the results from Ref.~\cite{Degrassi:2014sxa}.
  
The diboson ($WW$ and $WZ$)  and Higgstrahlung ($WH$ and $ZH$) fits use the data from Table IV of~\cite{Baglio:2020oqu} and we include NLO QCD. 
As shown in~\cite{Baglio:2020oqu}, the diboson and Higgstrahlung fits are extremely sensitive to whether the fit is performed at linear or quadratic order, with the $WZ$ contribution being particularly sensitive to the inclusion of NLO QCD effects. 
The diboson and VH data contribute to the fits predominantly through the high $p_T$ regime, and so it is important to
understand whether the EFT is valid in the regime where it is being applied.  
An extensive study was performed in Ref.~\cite{Baglio:2020oqu}, where it was shown that a large portion of the allowed region in the fits corresponds to a strongly coupled theory where higher dimension operators need to be included in future studies (see in particular, Fig.~7 of that paper). 
This is not true in the  fits to EWPO, where the fits fall within the weakly coupled regime. 
We further consider this question of EFT validity in Ref.~\cite {Baglio:2019uty} for the case of WZ production. 
In that study, we remove the high $m_{T}^{WZ}$ bins and show the effects on the fit.
Removing the bin with $m_{T}^{WZ}> 600~\textrm{GeV}$ from the fit, all points satisfy $m_{T}^{WZ} < \Lambda=1\,\textrm{TeV}$ and the EFT is a valid expansion.
In this case, the fits to the coefficients affecting the $WWZ$ vertex are degraded by roughly $30\%$. 

Finally, the Higgs predictions use the $80\,\textrm{fb}^{-1}~ 13\,\textrm{TeV}$ LHC data from ATLAS~\cite{Aad:2019mbh} and the $36-137\,\textrm{fb}^{-1}~ 13\,\textrm{TeV}$ LHC data from CMS~\cite{CMS:2020gsy}, along with the $8\,\textrm{TeV}$ data given in Tables 2 and 3 of~\cite{Ellis:2018gqa}.
The contribution to Higgs production and decay from $\Op_H$ occurs at loop order and is included following the prescription of Ref.~\cite{Degrassi:2016wml}. 
The identification of the observables with the SMEFT predictions is made using tree- level calculations (except for $C_{HG}$ and
$C_H$) and compared with the results of Refs.~\cite{ATLAS:2019dhi, Ellis:2018gqa}  and reasonable agreement is found.

In the following  subsections, we present results for our test models and discuss the use of the global EFT fits for extracting information about the underlying models.

\newpage
\subsection{Model Dependent Results}

\begin{figure}
\centering
\includegraphics[width=0.49\linewidth]{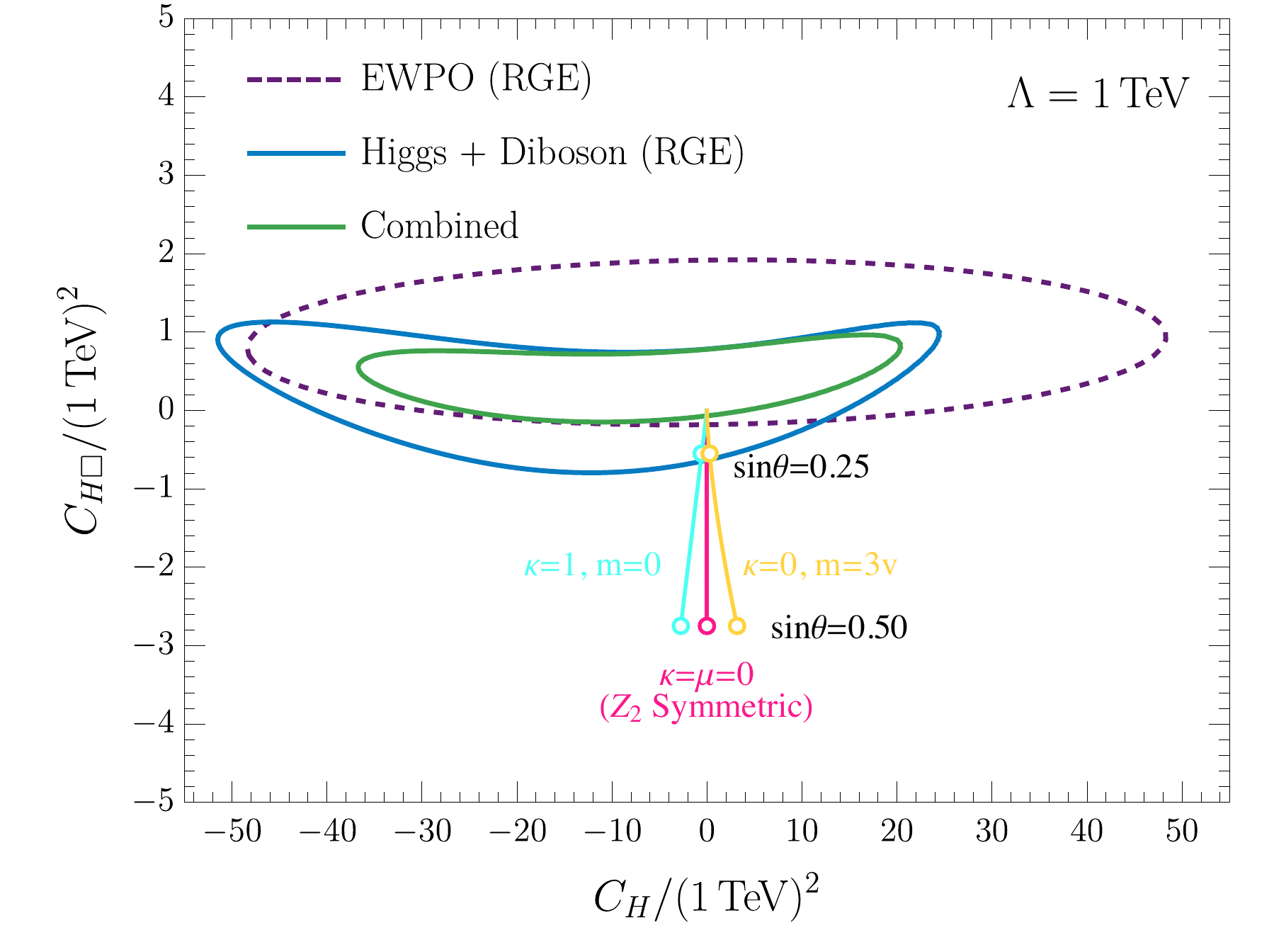}
\hfill
\includegraphics[width=0.49\linewidth]{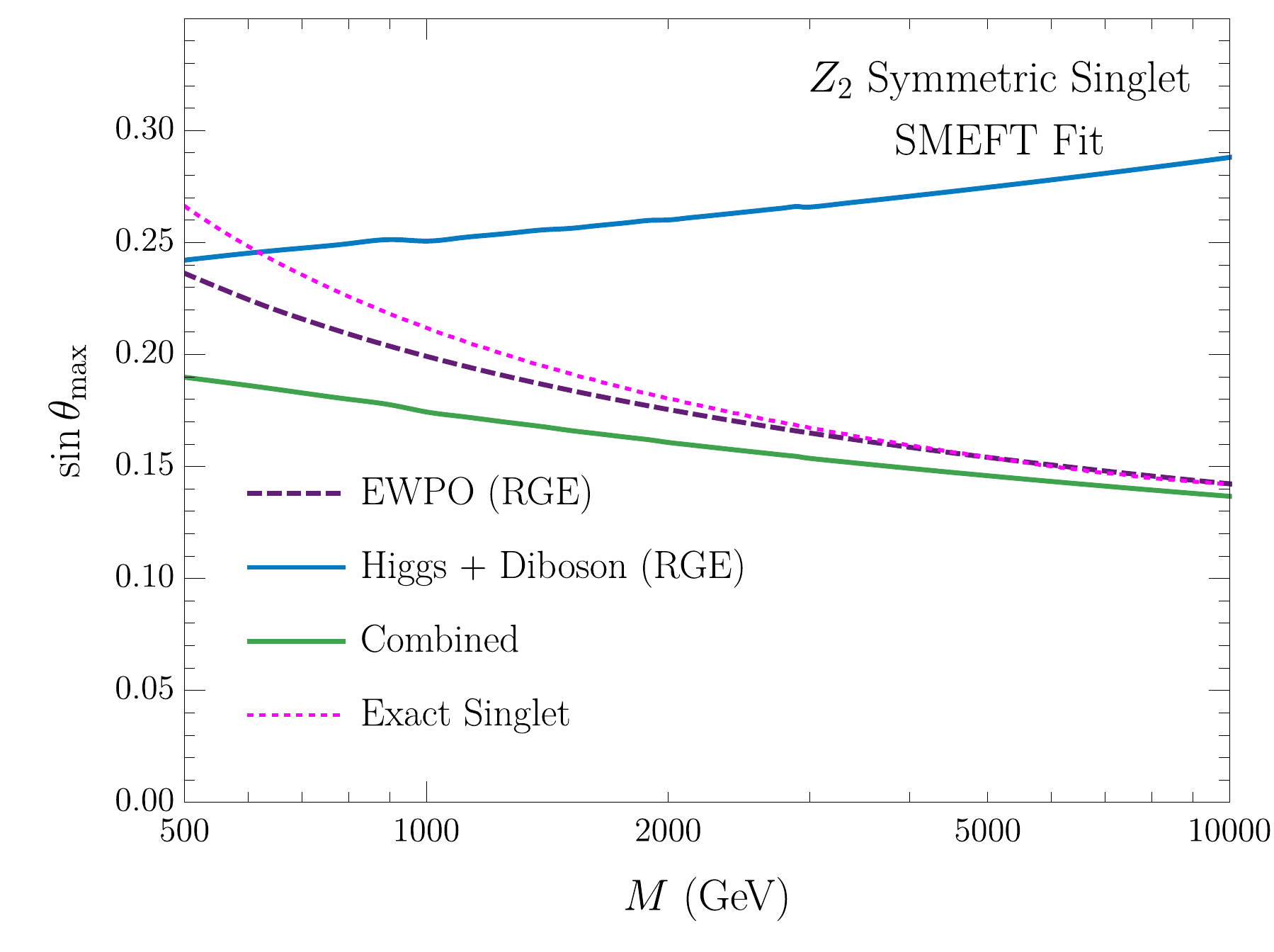}
\vskip -0.25cm
\caption{\label{fig:singfig}
LHS: $95\%$~C.L. limits on the Wilson coefficients $C_{H\square}$ and $C_H$ generated at the matching scale. 
The magenta, cyan, and yellow curves show combinations generated by the singlet model assuming a $Z_2$ symmetry, and with $\kappa = 1, m=0$, and $\kappa = 0, m = 3v$, respectively. The open circles indicate the point along the curve with $\sin\theta = 0.25, 0.50$.
RHS: The SMEFT coefficient limits reinterpreted in terms of a limit on the mixing angle for the $Z_2$ symmetric case as a function of the matching scale, $M$.
In both figures, the complete RGE of $C_{H\square}$ and $C_H$ from $\Lambda=1\,\textrm{TeV}$ to the weak scale is included.
}
\end{figure}

\subsubsection{Singlet Model}
\vskip -0.25cm

The singlet model generates the two operators, $\Op_{H\square}$ and $\Op_{H}$, at the matching scale $\Lambda$, which we take to be $M$, the mass of the heavier Higgs.
We first consider the results of a fit to arbitrary values of  the coefficients corresponding to these operators, assuming only that they are generated at the same scale, and taking all other operators to vanish. The results of this fit are shown on the LHS of Fig.~\ref{fig:singfig}, where the axes show the coefficients evaluated at the matching scale, $\Lambda = 1\,\mathrm{TeV}$.
$\Op_{H\square}$ leads to a shift in the Higgs-gauge couplings, and a universal change in the Higgs couplings to fermions, and these shifts are constrained by LHC data.
In addition, via Eq.~\ref{eq:regsum}, a non-zero $\Op_{H\square}$ at the matching scale generates the operators 
$\Op_{HD}$, $\Op_{Hq}^{(1)}$,~ and $\Op_{Hq}^{(3)}$ at the weak scale, yielding shifts to the EWPOs proportional to $\log(\Lambda / M_Z)$.
Numerically, the most important of these operators is $\Op_{HD}$, which generates the $T$ oblique parameter
\footnote{
The $S$ parameter depends on $C_{HWB}$ which is not generated from RGE in the singlet model.
As demonstrated in Ref.~\cite{Wells:2015cre}, in the Warsaw basis there are additional contributions from $4-$fermion operators at dimension-$6$ that  are needed to obtain a basis independent result for $S$.
It is also interesting to note from Ref.~\cite{deBlas:2017xtg}, that only a very small class of models generate $\Op_{HWB}$ from dimension-$6$ operators and none of the models considered here fall into this class.}.
The limits from EWPOs are shown in Fig.~\ref{fig:singfig} LHS as a purple dashed contour. 
As anticipated, the limit on $C_{H}$ is very weak~\cite{Kribs:2017znd,Degrassi:2014sxa}.
The constraint from the  combination of LHC Higgs and diboson data is shown as a solid blue line, and we see that measurements of the Higgs couplings provide a bound on $C_{H\square}$ of the same order as the EWPO fit. The combination of LHC data and EWPOs is shown as the solid green curve.
In single Higgs data, the operators $\Op_{H\square}$ and $\Op_{H}$ do not generate any momentum dependence, but 
$\Op_{H\square}$ produces momentum-dependent effects in di-Higgs data which could potentially be of use for the discrimination between models~\cite{Brehmer:2015rna, Capozi:2019xsi, DiVita:2017eyz}.

While we consider first general values of $C_{H}$ and $C_{H\square}$, the singlet model generates only a subset of these coefficients.
In the $Z_2$ symmetric case, $C_{H} = 0$, and $C_{H\square}$ is always less than zero (see Appendix~\ref{sec:scalars}), so this class of models generates a vertical ray emanating from the origin, shown as a magenta curve in Fig.~\ref{fig:singfig} LHS, with the open circles indicating values of the mixing angle, $\sin\theta = 0.25, 0.50$.
  The global fit clearly excludes values of $\sin\theta$ above $\sim 0.25$.  
Furthermore, in order for the SMEFT to be valid, the coefficients cannot be too large.  If we require ${v^2\over\Lambda^2} C_{H\square}\lsim {1\over 2}$, this would correspond to the requirement that
$\sin\theta \lsim 0.7$, implying that our global fit limits are well within the range of validity of the global fit. 
 Limits on the mixing angle from unitarity and perturbativity of the couplings are considerably weaker than those from the global fits~\cite{Robens:2015gla}.

In the absence of the $Z_2$ symmetry, the additional couplings lead to more general combinations of $C_H$ and $C_{H\square}$ at the matching scale, and we show two such combinations, varying the mixing angle, as yellow and cyan rays in Fig.~\ref{fig:singfig} LHS. The values chosen saturate the vacuum stability bounds, Eq.~\ref{eq:vacans},  for $\kappa$ and $m = 0$, respectively. We see that only a small slice of the 2 dimensional SMEFT parameter space at the matching scale can be consistently generated in the singlet model, and that this slice covers only a small part of the allowed range.

With the other coefficients held fixed, the SMEFT limits can be translated into limits on the physical parameter, $\sin\theta$ (defined in Appendix~\ref{sec:scalars}).
In the $Z_2$ symmetric case, it is apparent from the LHS that $\sin\theta \lesssim 0.25$ for $\Lambda = 1\,\mathrm{TeV}$.
More generally, we can interpret limits on the coefficients as limits on the largest allowed mixing angle as a function of the scale, $\Lambda$.
For the $Z_2$ symmetric case, $v^2 C_{H\square} / \Lambda^2 = -{1\over 2}\tan^2\theta$, so the only scale dependence is the weak logarithmic dependence from RG evolution, and this is shown on the RHS of Fig.~\ref{fig:singfig}.
The EWPO limits obtained by fitting the SMEFT coefficients are quite similar to those extracted directly from the EWPOs in the full singlet model~\cite{Falkowski:2015iwa, Lopez-Val:2014jva} 
as shown by the magenta curve on the RHS  of Fig.~\ref{fig:singfig},
although there is a roughly $20\%$ difference that can be attributed to the non-logarithmic terms in the complete
model calculation~\cite{Dawson:2009yx}.
While $C_{HD}$  also appears in the LHC Higgs and diboson data, the bounds on $C_{H\square}$ arise at tree level, and there are thus different effects in the scaling on the RHS of Fig.~\ref{fig:singfig}.

\subsubsection{2HDM}
\vskip -0.25cm

\begin{figure}
 \centering
\includegraphics[width=0.49\linewidth]{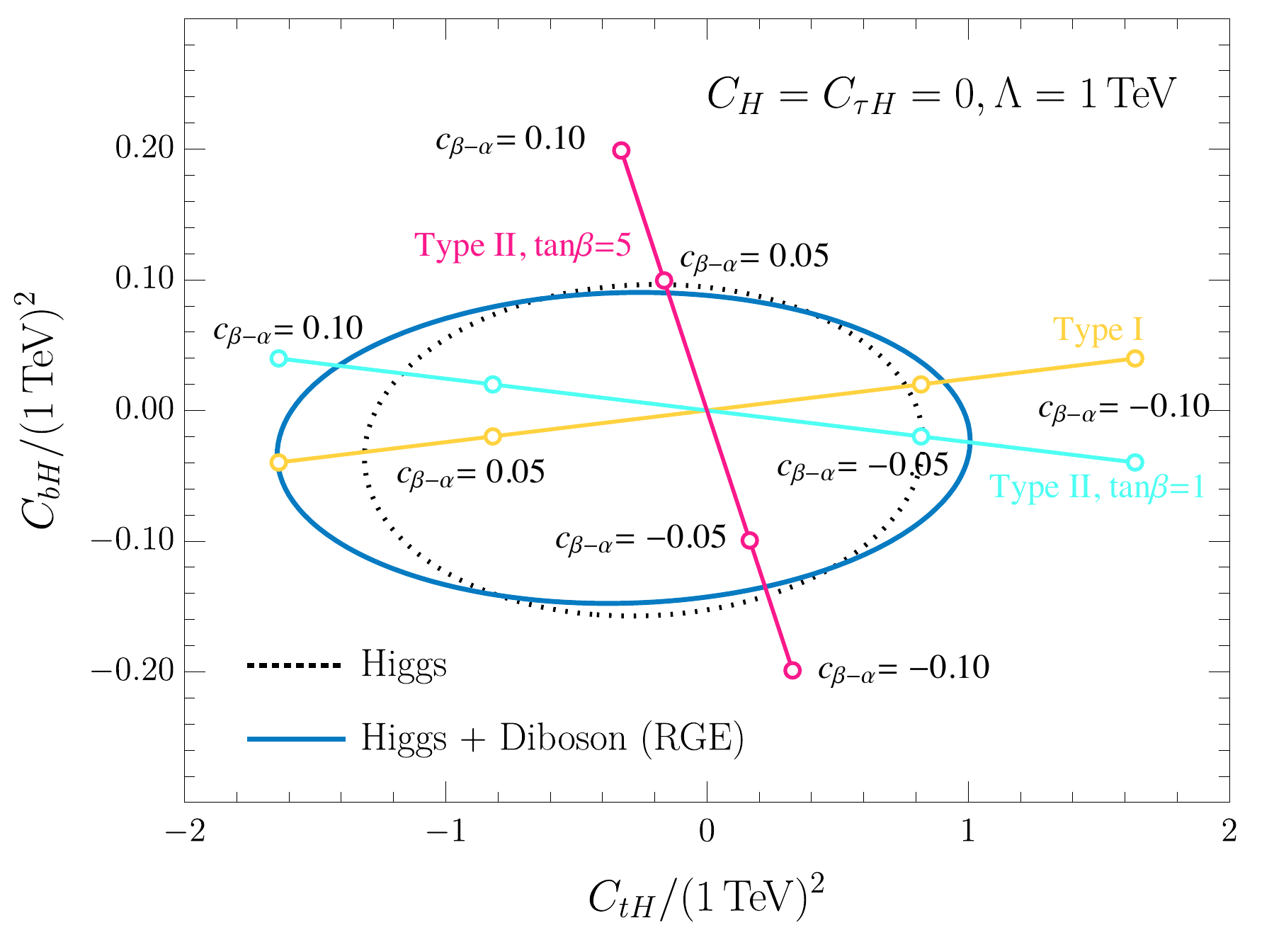}
\hfill
\includegraphics[width=0.49\linewidth]{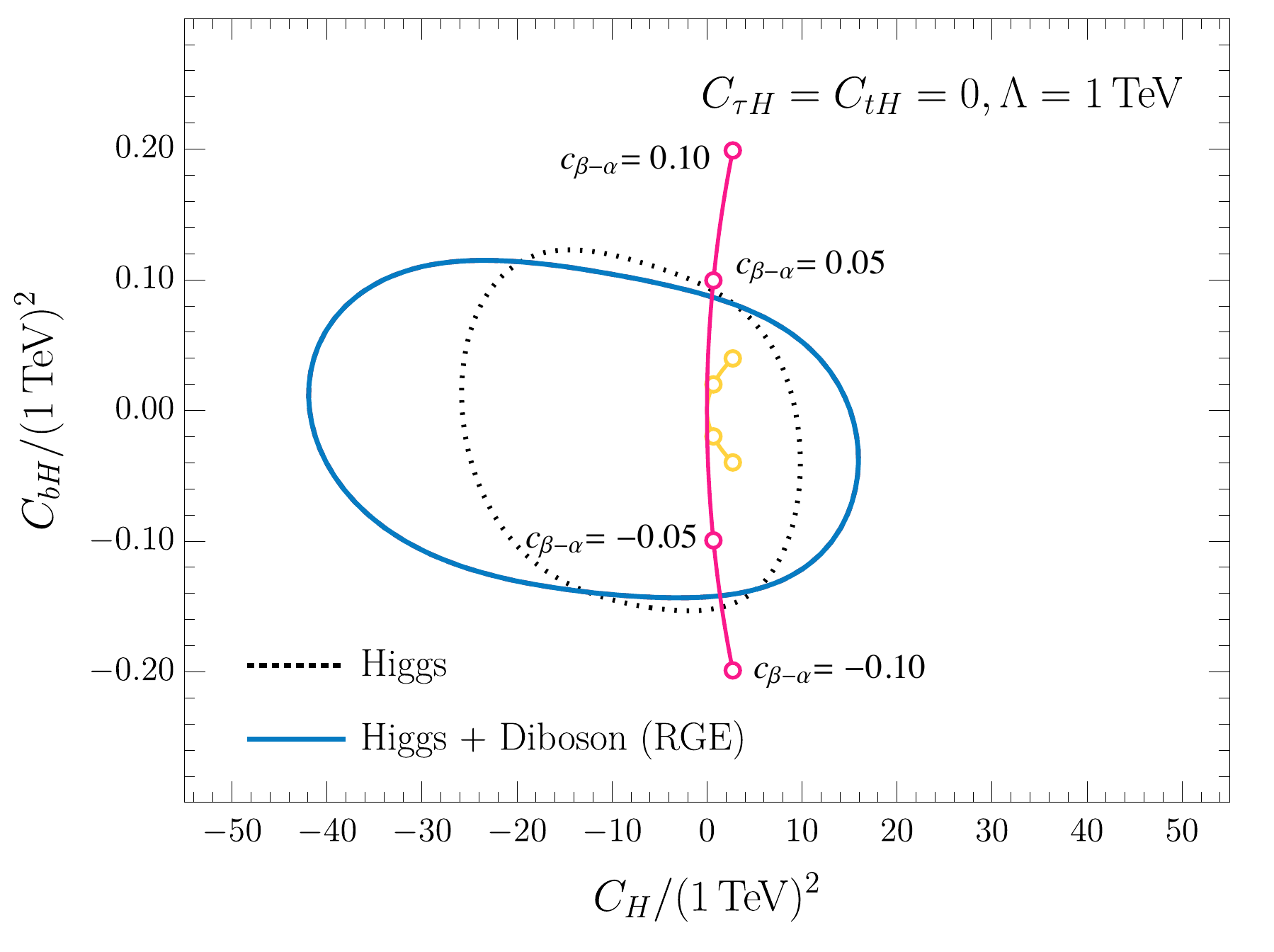}
\vskip -0.25cm
\caption{$95\%$ confidence level limits in the 2HDM from LHC data.  All other coefficients are set to zero. 
The yellow curve corresponds to a Type I model with $\tan\beta = 1$, while the cyan and magenta curves correspond to Type II models with $\tan\beta = 1, 5$ respectively.
Open circles along the curve indicate points with $\cos(\beta - \alpha) = \pm 0.05, 0.10$. 
The coefficients on the axes are evaluated at the matching scale $\Lambda$ and the curves labelled ``Higgs'' do not
include the RGE of the operators.
 }
\label{fg:2hdmfig}
\end{figure}

The decoupling limit of the 2HDM has been extensively discussed in the literature~\cite{Gunion:2002zf, Haber:2015pua}, but here we revisit the question of what information is in the SMEFT fits in the limit where the new scalars are too heavy to be observed.
In Appendix~\ref{sec:2hdm}, we see that the 2HDM generates $\Op_H$, $\Op_{uH}$, $\Op_{dH}$ and $\Op_{eH}$ at tree-level, assuming a small deviation from the alignment limit, $\mid \cos(\beta - \alpha)\mid \ll 1$.
In general, these operators can have arbitrary coefficients for each generation, but to avoid flavor-changing neutral currents, they are usually assumed to be proportional to the SM Yukawa matrices, so --- in the limit that only the third generation Yukawas are non-zero --- we can consider only the $33$ components of these fermionic operators, and label them $\Op_{tH}$, $\Op_{bH}$, and $\Op_{\tau H}$
\footnote{Consistent models with large Yukawa couplings to the first and second generations can also be constructed~\cite{Bishara:2016jga,Alasfar:2019pmn,Egana-Ugrinovic:2019dqu}.}.

\begin{figure}
 \centering
\includegraphics[width=0.49\linewidth]{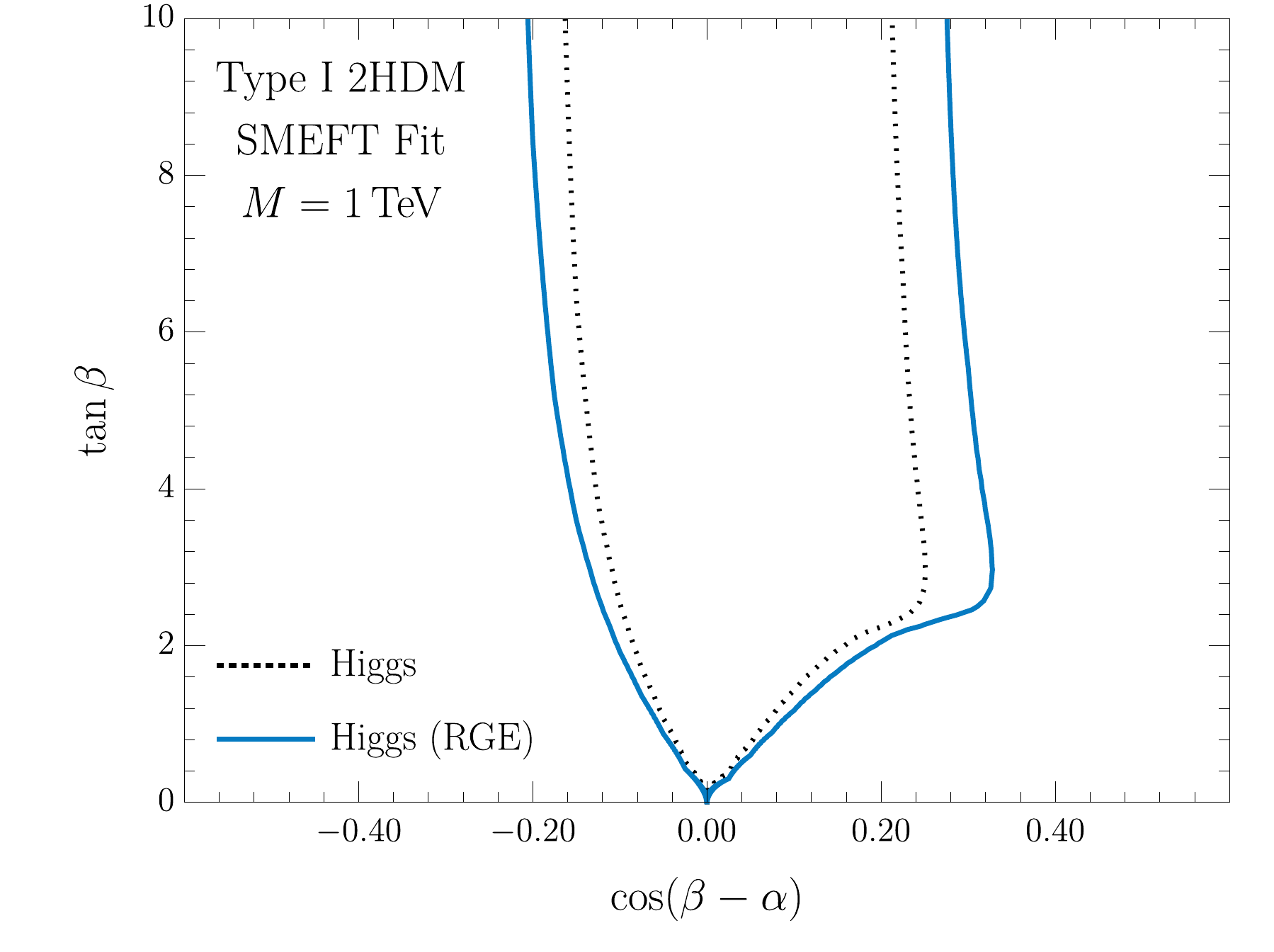}
\hfill
\includegraphics[width=0.49\linewidth]{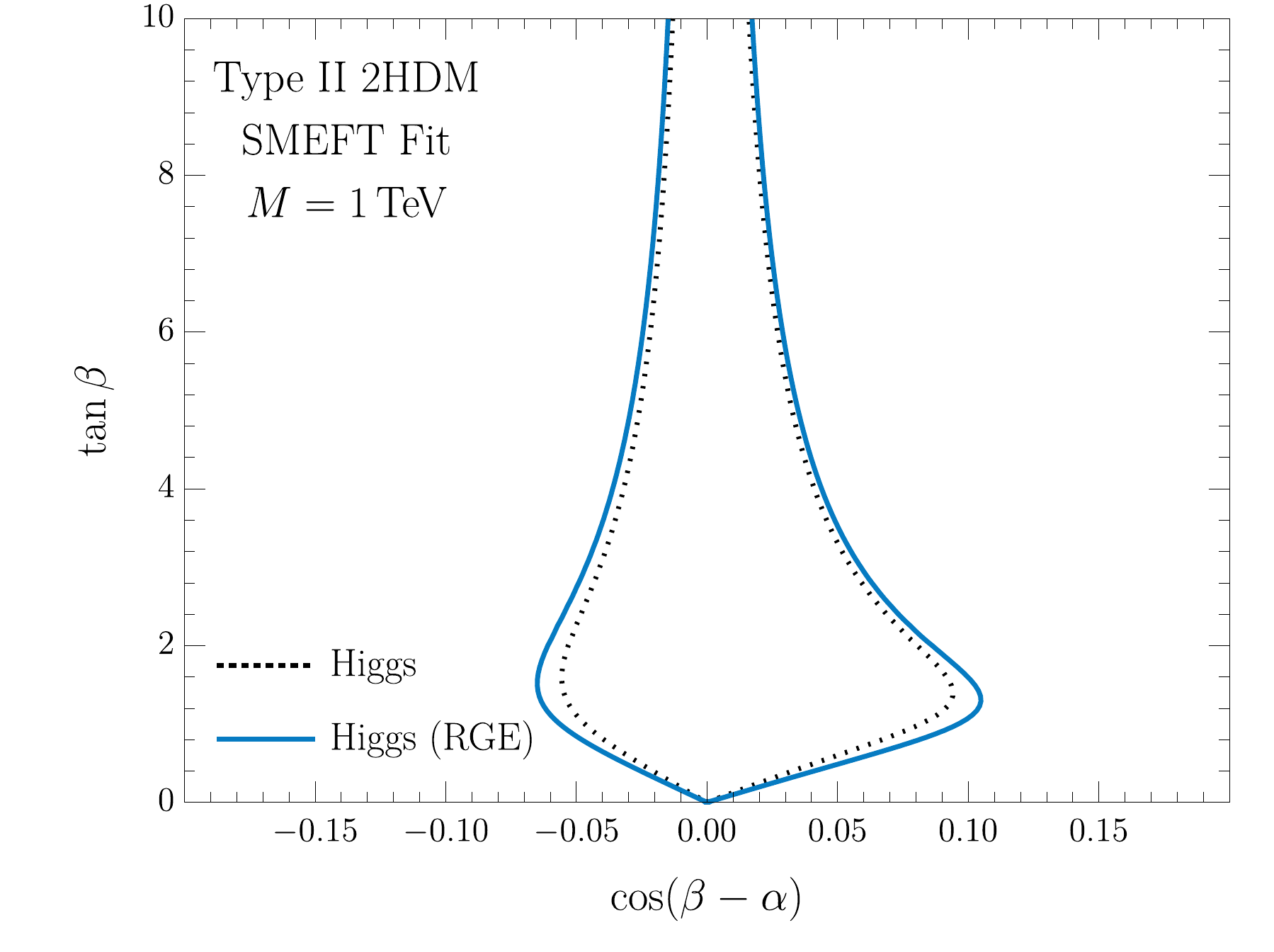}
\vskip -0.25cm
\caption{
$95\%$ C.L. limits using the SMEFT fit  on the Type I and Type II 2HDM from LHC data in the $\cos(\beta - \alpha)$ vs. $\tan\beta$ plane. 
 The curves labelled ``Higgs'' do not
include the RGE of the operators. Note that our results are formally only valid in the $\mid \cos(\beta-\alpha)\mid \ll1$ limit.
 }
\label{fg:2hdmphysfig}
\end{figure}

None of these operators contribute to the EWPOs at tree level, nor do they generate any of the operators in Eq.~\ref{eq:regsum} at
 leading logarithm, so we consider only LHC constraints.
In Fig.~\ref{fg:2hdmfig}, we consider the limits in the $C_{tH}$--$C_{bH}$ plane (LHS) and the $C_H$--$C_{bH}$ plane (RHS), in each case setting all other operators to zero.
The dashed line shows a fit to Higgs data with the scale-dependence of the operators neglected, while the solid blue curves show the results of a fit where the operators are generated at $\Lambda = 1\,\mathrm{TeV}$ and then RG evolved down to $M_Z$.
On the LHS, we see that the RGE has a minor effect, running the Yukawa-like operators to smaller values in the UV, and resulting in slightly weaker bounds on the coefficients.
As in the singlet model case, we see that $C_H$ is only very weakly constrained, and note that the effect of the RGE on
the RHS is significant in this case.

In both panels of Fig.~\ref{fg:2hdmfig}, we show the patterns of coefficients generated by different choices of the Glashow-Weinberg conditions~\cite{Glashow:1976nt} for the proportionality constants in the fermion-Higgs operators.
The yellow curve corresponds to a Type I model with $\tan\beta = 1$, while the cyan and magenta curves correspond to Type II models with $\tan\beta = 1, 5$ respectively. Note that the $\tan\beta = 1$ curves for the Type I and Type II models are indistinguishable on the RHS.
The distances along these curves correspond to different values of the alignment parameter, and we indicate values of $\cos(\beta - \alpha) = \pm 0.05$ and $\pm 0.1$ by open circles.
We see that most of the allowed parameter space in the $C_{tH}$ vs. $C_{bH}$ plane can be generated by considering different classes of 2HDMs with different values of $\cos(\beta - \alpha)$ and $\tan\beta$.

The fit to SMEFT coefficients is re-interpreted in terms of the parameters of the 2HDM in Fig.~\ref{fg:2hdmphysfig}, for both the Type I and Type II models  near the alignment limit with $M = 1\,\mathrm{TeV}$.
These fits show good agreement with the fits in the full, UV complete 2HDM~\cite{Belusca-Maito:2016dqe}.
In the Type I 2HDM, the effects of the RGE reduce the value of $C_{tH}$ when scaling from $M_Z$ to $\Lambda$  as observed in Fig.~\ref{fg:2hdmfig} and are manifest in the difference between the solid and dashed line in Fig.~\ref{fg:2hdmphysfig} as well.

\subsubsection{Heavy Colored particles in Loops: $T$ VLQ}
\vskip -0.25cm

\begin{figure}
\centering
\includegraphics[width=0.49\linewidth]{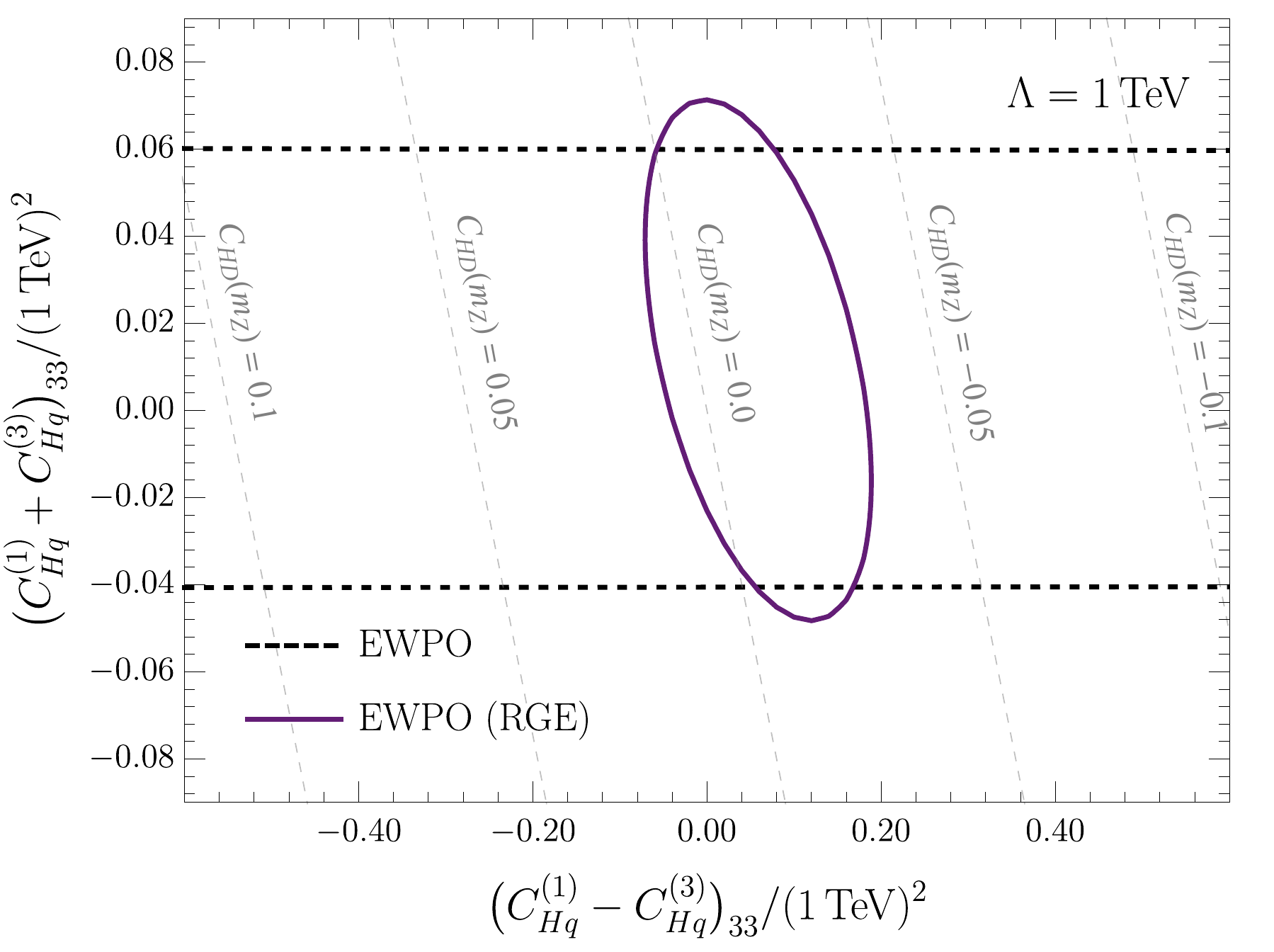}
\hfill
\includegraphics[width=0.49\linewidth]{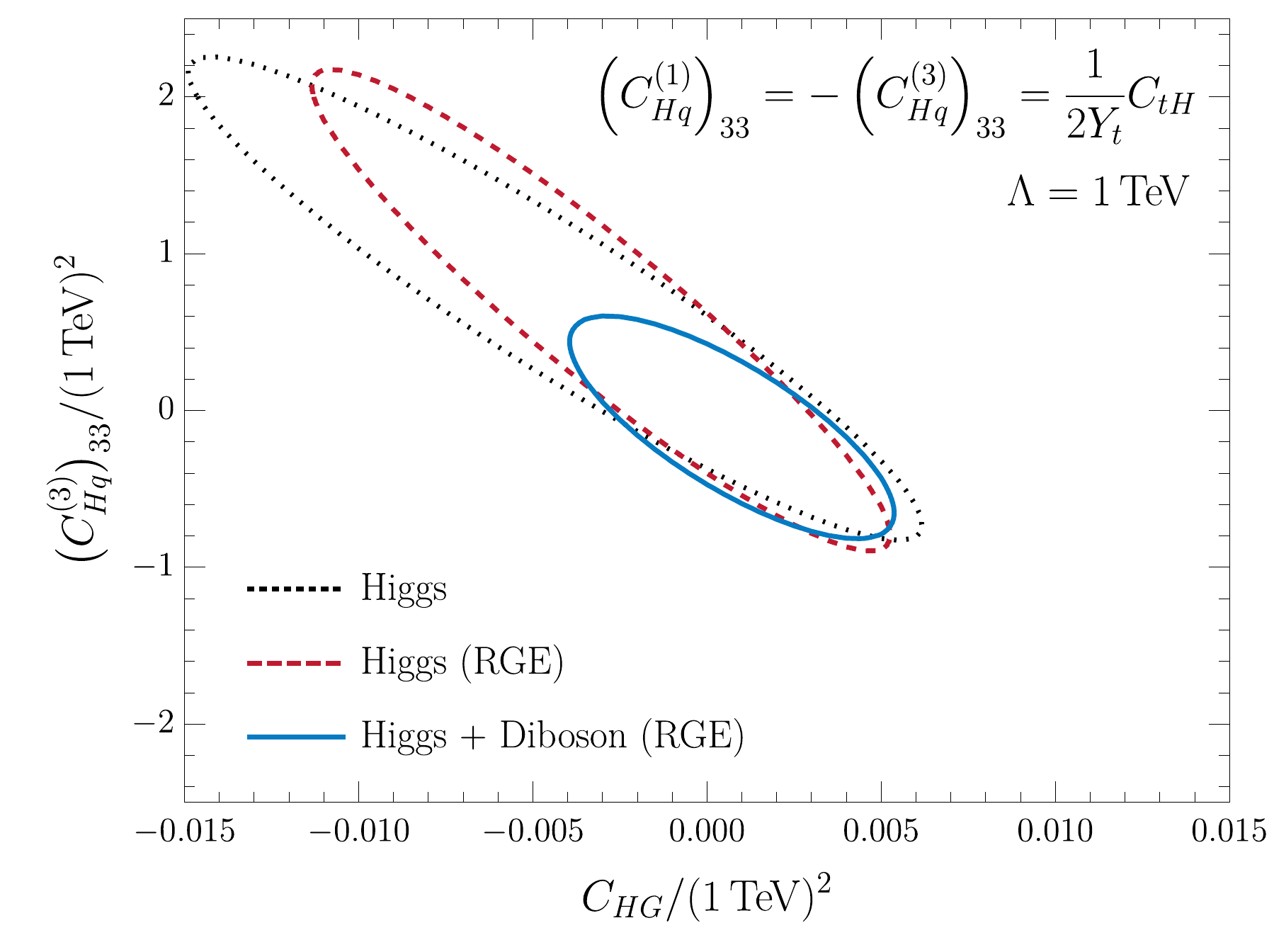}
\vskip -0.25cm
\caption{
LHS: $95\%$ confidence level limits from EWPOs (LHS) and 
RHS: Limits from Higgs and diboson data with SMEFT coefficients in the pattern generated in the $T$ VLQ model.  The curve labelled ``Higgs'' does not include RGE effects.
The coefficients on the axes are evaluated at $\Lambda=1\,\mathrm{TeV}$.
}
\label{fg:tvlqfig}
\end{figure}

The case of a heavy $Q = {2\over 3}$ vector-like quark coupling only to the third generation leads to the operators $\Op_{tH}$, $\big(\Op_{Hq}^{(1)}\big)_{33}$ and $\big(\Op_{Hq}^{(3)}\big)_{33}$ at tree-level, as well as $\Op_{HG}$ at one loop.
Importantly, we note that only the third generation operators appear, so fits treating these fermion operators universally are impossible to interpret in the context of these models.

We first consider the constraints on only $\big(C_{Hq}^{(1)}\big)_{33}$ and $\big(C_{Hq}^{(3)}\big)_{33}$ from EWPOs.
The dominant effect at tree-level is in the $Z\to b\bar{b}$ decay, which depends only on the combination $\big(C_{Hq}^{(1)} + C_{Hq}^{(3)}\big)_{33}$. This is indicated by the horizontal dashed black lines in Fig.~\ref{fg:tvlqfig} LHS.
At leading logarithm, this degeneracy is broken  and both $\big(C_{Hq}^{(1)}\big)_{33}$ and $\big(C_{Hq}^{(3)}\big)_{33}$ contribute to the anomalous dimension of $\Op_{HD}$. Lines of constant $C_{HD}(M_Z)$ generated from the RGE are shown as dashed gray lines in Fig.~\ref{fg:tvlqfig} LHS, and the resulting limit from the EWPOs including the RGEs, shown as a solid purple curve, clearly follows this shape.
It is  clear that to draw any conclusions about this model from the EWPOs including both the RGE effects and the non-universality of the fermion operators is crucial.

\begin{figure}
\centering
\includegraphics[width=0.49\linewidth]{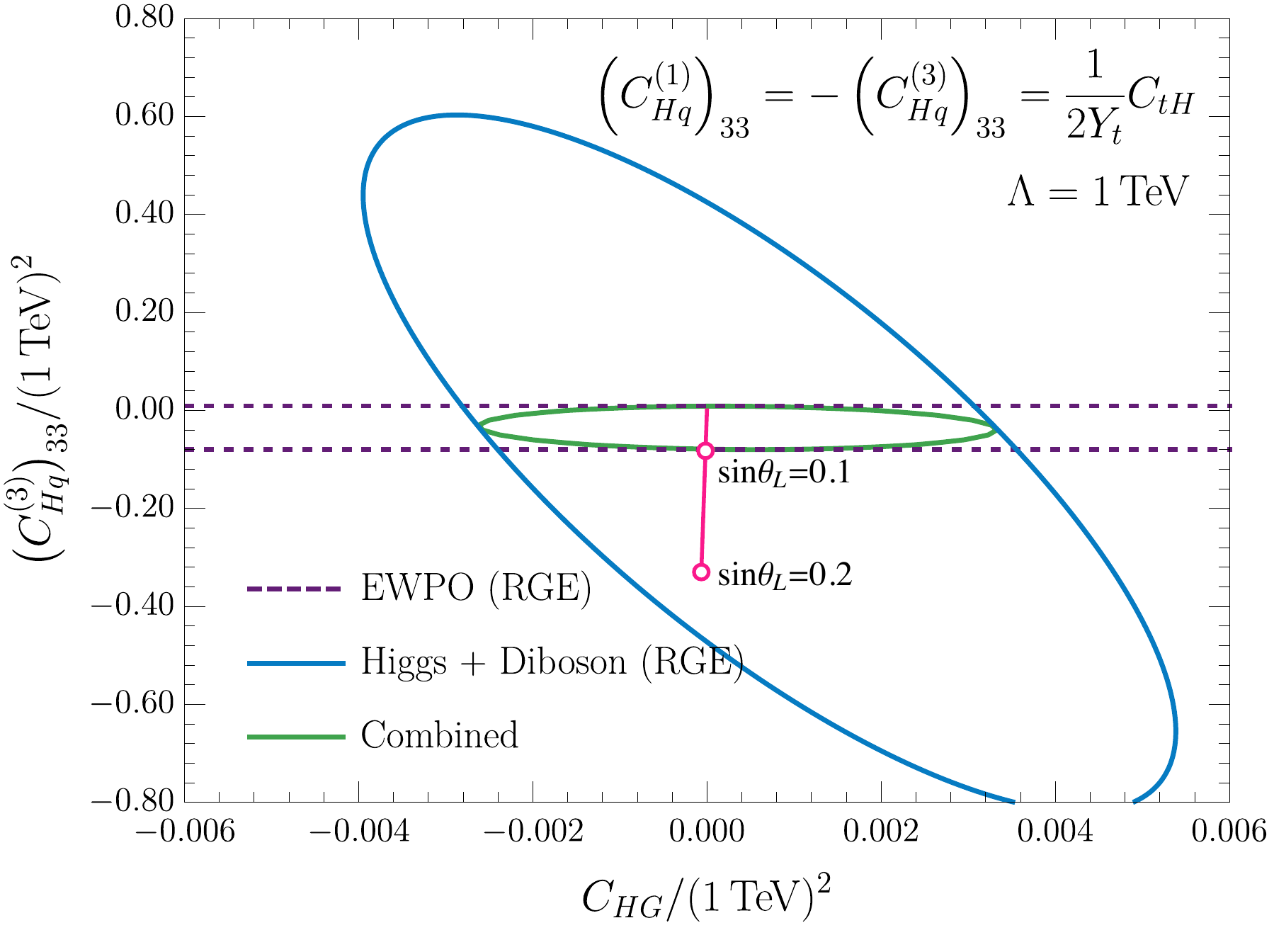}
\hfill
\includegraphics[width=0.49\linewidth]{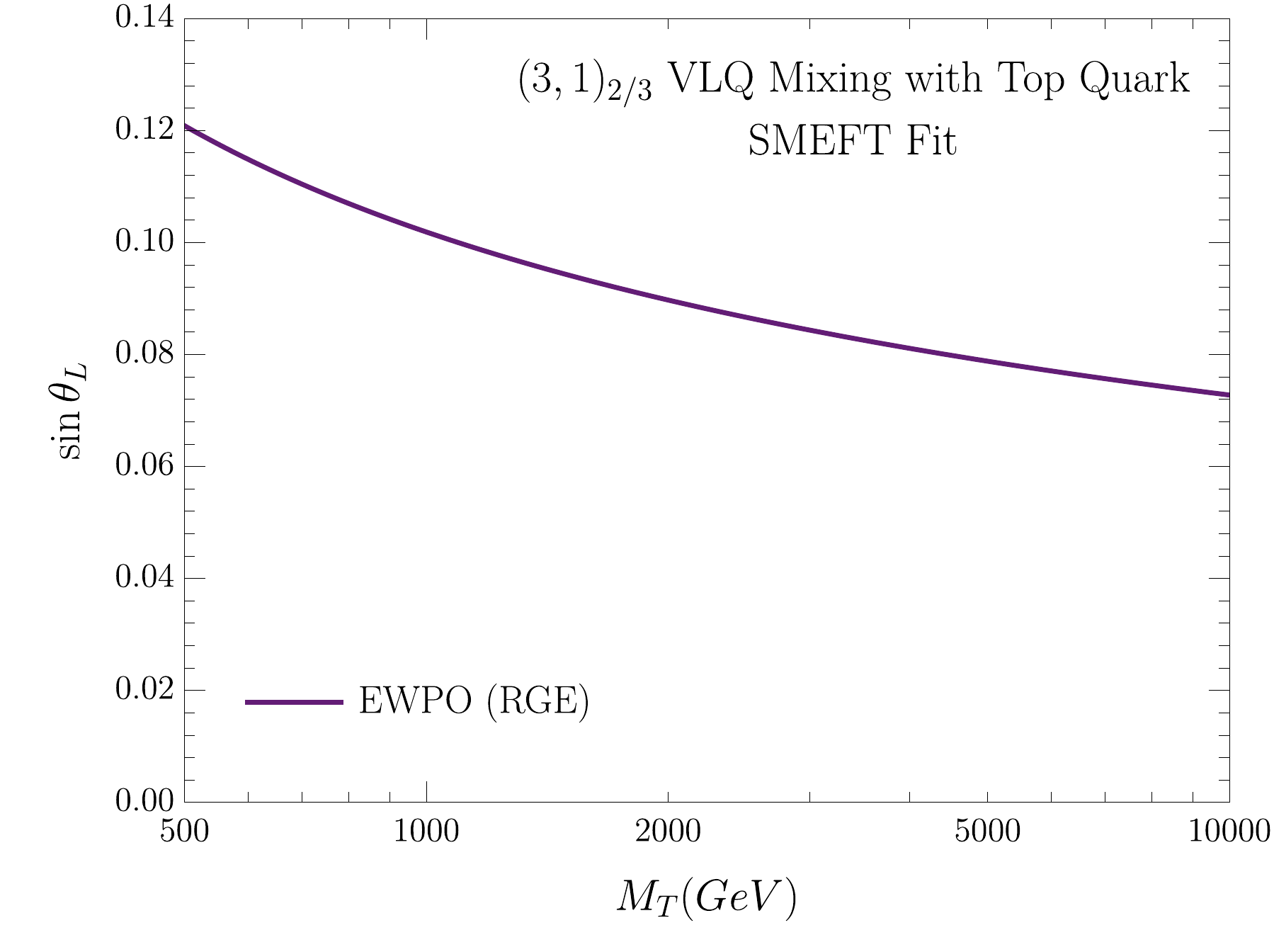}
\vskip -0.25cm
\caption{LHS: Global fit limits with coefficients in the pattern generated in the $T$ VLQ.
The green line is the combination of EWPOs, Higgs, and diboson data, with all RGE effects included.
The magenta line demonstrates  the dependence on the mixing angle corresponding to the SMEFT fits with $s_L^t\equiv \sin \theta_L$.
The coefficients on the axes are evaluated at $\Lambda=1\,\textrm{TeV}$.
RHS:  A translation of the limits of the LHS to the parameters of the $T$ VLQ model.
}
\label{fg:tvlqmodel}
\end{figure}

On the RHS, we consider the limits from Higgs and diboson data. As is apparent from Table~\ref{tab:sumlag}, in the context of the $T$ VLQ, the Wilson coefficients that are generated all come with a fixed pattern, since there is only one independent physical parameter.
To illustrate the importance of this, we thus fix all the relationships between the coefficients except for $C_{HG}$ and plot in the $\big(C_{Hq}^{(3)}\big)_{33} = - \big(C_{Hq}^{(1)}\big)_{33} = - C_{tH}/Y_t$ vs. $C_{HG}$ plane.
The resulting limits are highly correlated, and we see that including the effects of the RGE on the fit with only Higgs data (going from dotted black to dashed red lines) has a mild effect on the fits.
Most of the limit in the vertical direction, however, comes from including the diboson data, 
and this dependence arises because  the RGE generates $\Op_{H\square}$, which is well constrained by $VH$ and $VV$ fits~\cite{Baglio:2020oqu,Brehmer:2019gmn}.

Fig.~\ref{fg:tvlqmodel} LHS uses the same set of correlations for the Wilson coefficients, and compares the EWPOs constraint (purple, dashed) with the Higgs plus  diboson constraint (blue, solid), all with the RGE included. The EWPOs still set a superior limit on the fermionic operators from the RGE induced $\Op_{HD}$, while the Higgs data sets the bound on $\Op_{HG}$. The combined limit is shown in solid green.
In magenta, we show the full prediction of the model, with $\Op_{HG}$ related to the other operators via Eq.~\ref{eq:vlqt_chg} and open circles indicating a mixing angle of $\sin\theta_L\equiv s_L^t = 0.1, 0.2$. 
As noted above, the $T$ VLQ generates only a 1- dimensional slice of the parameter space, and while a small value of $C_{HG}$ can be generated, the values corresponding to the model are still nearly orthogonal to the EWPO constraints so that the LHC data is less important in setting a bound on the physical parameter. 
It is clear that for $M_T = 1\,\textrm{TeV}$, $s^t_L$ is restricted to $< 0.1$.
The RHS of Fig~\ref{fg:tvlqmodel} interprets the SMEFT results as limits on this mixing angle as a function of the $T$ VLQ mass. 
As in our other examples, the dependence on the scale is only logarithmic, from the RG evolution.

\begin{figure}[t]
\centering
\includegraphics[width=0.49\linewidth]{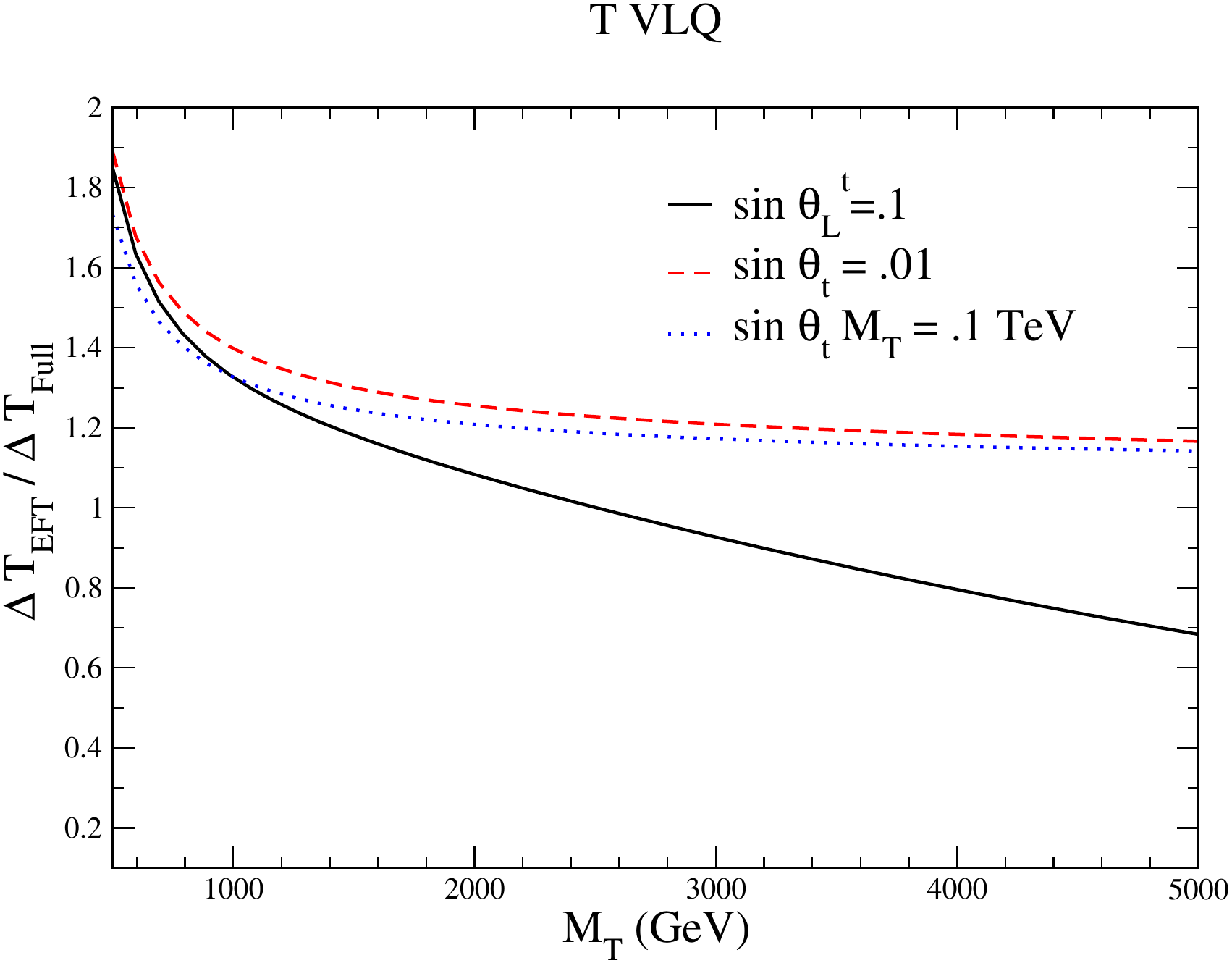}
\vskip -0.25cm
\caption{The ratio of the oblique parameter $\Delta T$ in the SMEFT limit with RGE to the result in the full $T$ VLQ model for
various values of the mixing parameter.  
}
\label{fg:tvlq_dt}
\end{figure}

 It is of interest to consider how well the global fit limits using the SMEFT coefficients reproduce the constraints found in the complete $T$ VLQ model.  A simple estimate can can found from considering the  oblique parameter $\Delta T$.  The $T$ VLQ result is 
given in Eq. \ref{eq:tuv}, while the SMEFT limit is derived in Eq. \ref{eq:teft} of Appendix \ref{sec:colored}.  The SMEFT construction reproduces the $\log({M_T\over m_t})$ terms, but not the constant terms,  in $\Delta T$.
The  ratio of the $\Delta T$ parameter in the SMEFT limit to the full result is shown in Fig. \ref{fg:tvlq_dt}.  
Even for quite small values of the mixing parameter, there is  a roughly $\sim 20~\%$ difference between the results obtained in the SMEFT  with tree level matching and one-loop RG running, as compared to the full theory for a fixed  value of  $\sin\theta_L^t$.
However, in order for the $T$ VLQ model to remain weakly coupled, the parameters of the Lagrangian need to scale
as $\sin\theta_L^t\sim {1\over M_T}$~\cite{Dawson:2012di}. This limit is also shown in the figure, and again the SMEFT limits differ by $\sim 20~\%$ from those of 
the full theory.

\subsubsection{Heavy Colored particles in Loops: $(T\,B)$ VLQ}
\vskip -0.25cm

\begin{figure}
\centering
\includegraphics[width=0.49\linewidth]{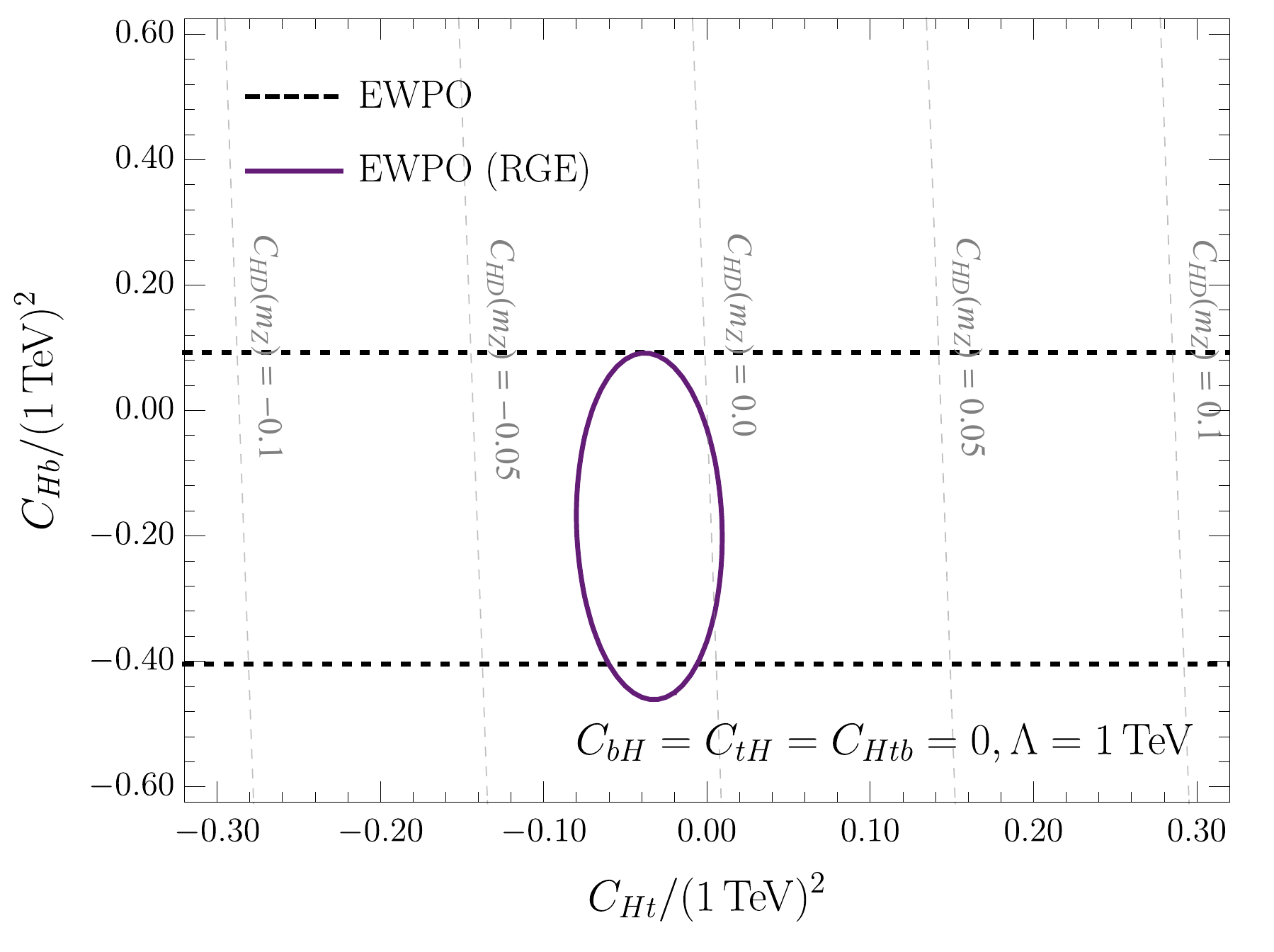}
\hfill
\includegraphics[width=0.49\linewidth]{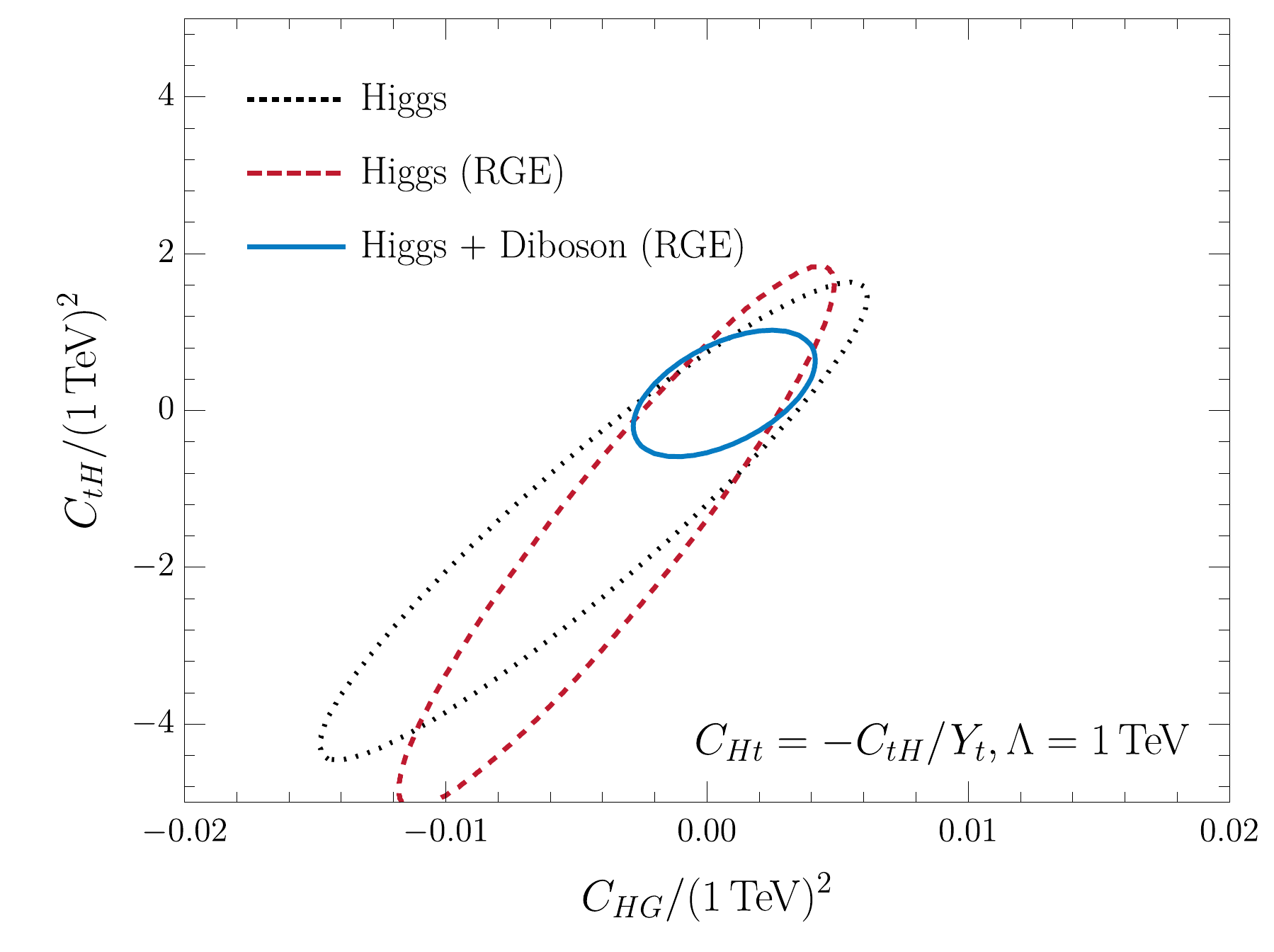}
\vskip -0.25cm
\caption{LHS:  $95\%$ confidence level  limits on $C_{Ht}$ and $C_{Hb}$ when all other coefficients are set to zero from
EWPOs.  RHS: Same, but including the RG evolution from the matching to the weak scale.  The coefficients
on the axes are evaluated at $\Lambda=1\,\textrm{TeV}$.
}
\label{fg:tbvlqmodelone}
\end{figure}

In the $(T\, B)$ VLQ model, the operators $\Op_{Ht},~ \Op_{Hb},~ \Op_{tH},~ \Op_{bH}$ and $\Op_{Htb}$ are generated at tree level, and we emphasize again that only the third generation fermion operators appear.
As in the $T$ VLQ model, the only tree-level contribution to EWPOs is through the $Z\to b\bar{b}$ decay, which now directly constrains $C_{Hb}$. 
This is shown as the dashed black lines on the LHS of Fig.~\ref{fg:tbvlqmodelone}, where we show the SMEFT limits in the $C_{Hb}$ vs. $C_{Ht}$ plane, with  the coefficients of all other operators fixed to zero. 
At tree level, the EWPOs are independent of $C_{Ht}$, but both $C_{Hb}$ and $C_{Ht}$ generate $C_{HD}$ via the RGE at leading
 logarithm. The contours of constant $C_{HD}$ (with the slope fixed by the relative values of $Y_t$ and $Y_b$, as expected from Eq.~\ref{eq:regsum}) are shown in dashed gray, and we again see that the EWPO limit including the RGEs (purple, solid) follows this pattern.

The limits coming from Higgs data and diboson data are illustrated on the RHS of Fig.~\ref{fg:tbvlqmodelone}.
Here, we include the operators $\Op_{tH}$, $\Op_{Ht}$ and $\Op_{HG}$ at the matching scale, with the
coefficients of other operators taken to be zero, and then project onto the plane where $C_{tH} = -Y_t\,C_{Ht}$, as predicted by the $(T\,B)$ VLQ.
Similarly to the $T$ VLQ case, the importance of the diboson data is apparent when the RGE of the  coefficients from the matching scale  to the weak scale is included, as a nonzero $C_{Ht}$ at the matching scale leads to values of $C_{HD}$ at the weak scale that are well constrained by diboson data. 
A similar story pervades three of our example models: if a $C_{HD}$ or $C_{H\square}$ appear at leading logarithm in the anomalous dimensions for the operators  generated at the matching scale, the diboson data can place important constraints that may not be apparent in the tree-level matching.

\begin{figure}
\centering
\includegraphics[width=0.49\linewidth]{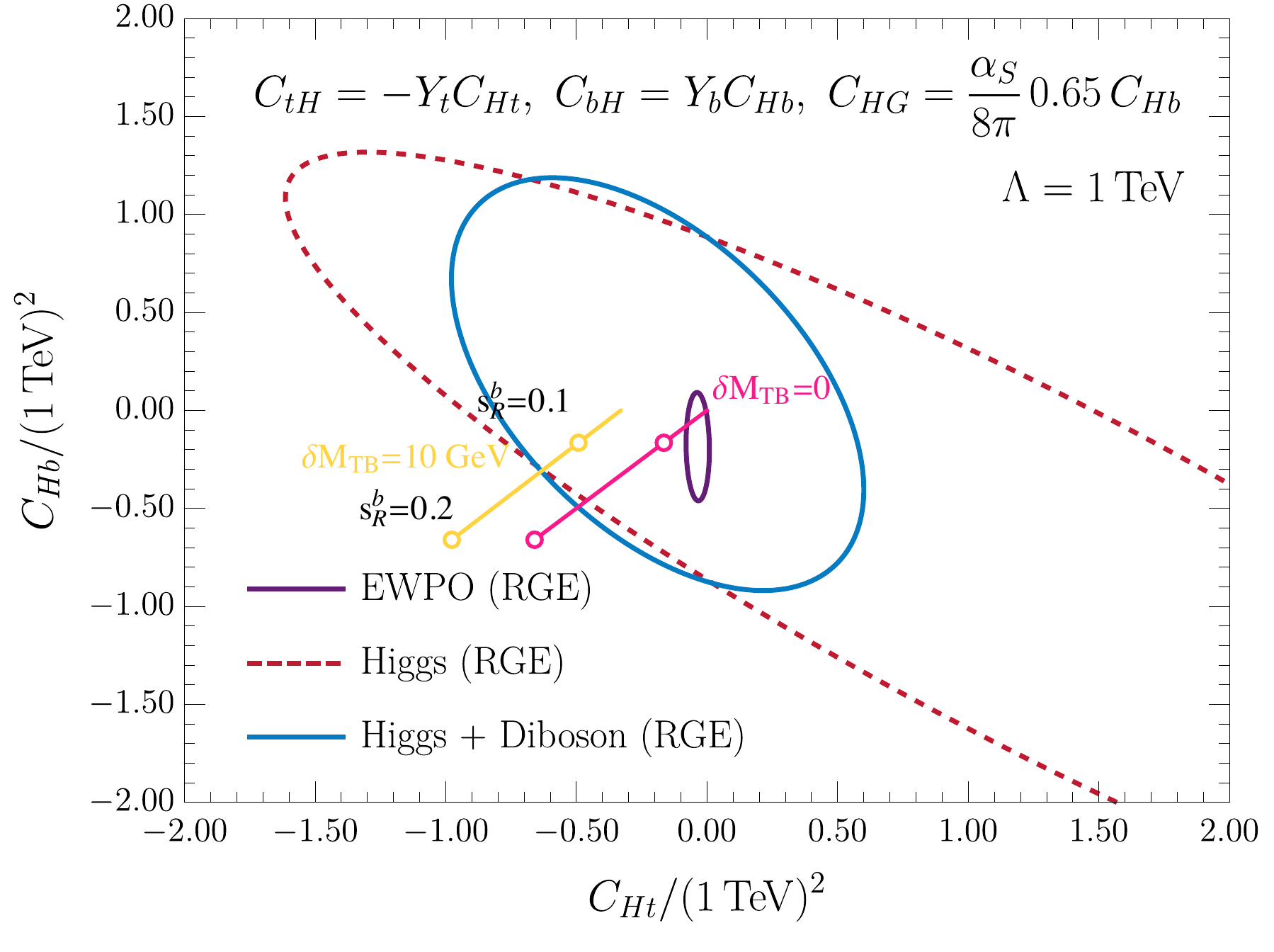}
\hfill
\includegraphics[width=0.49\linewidth]{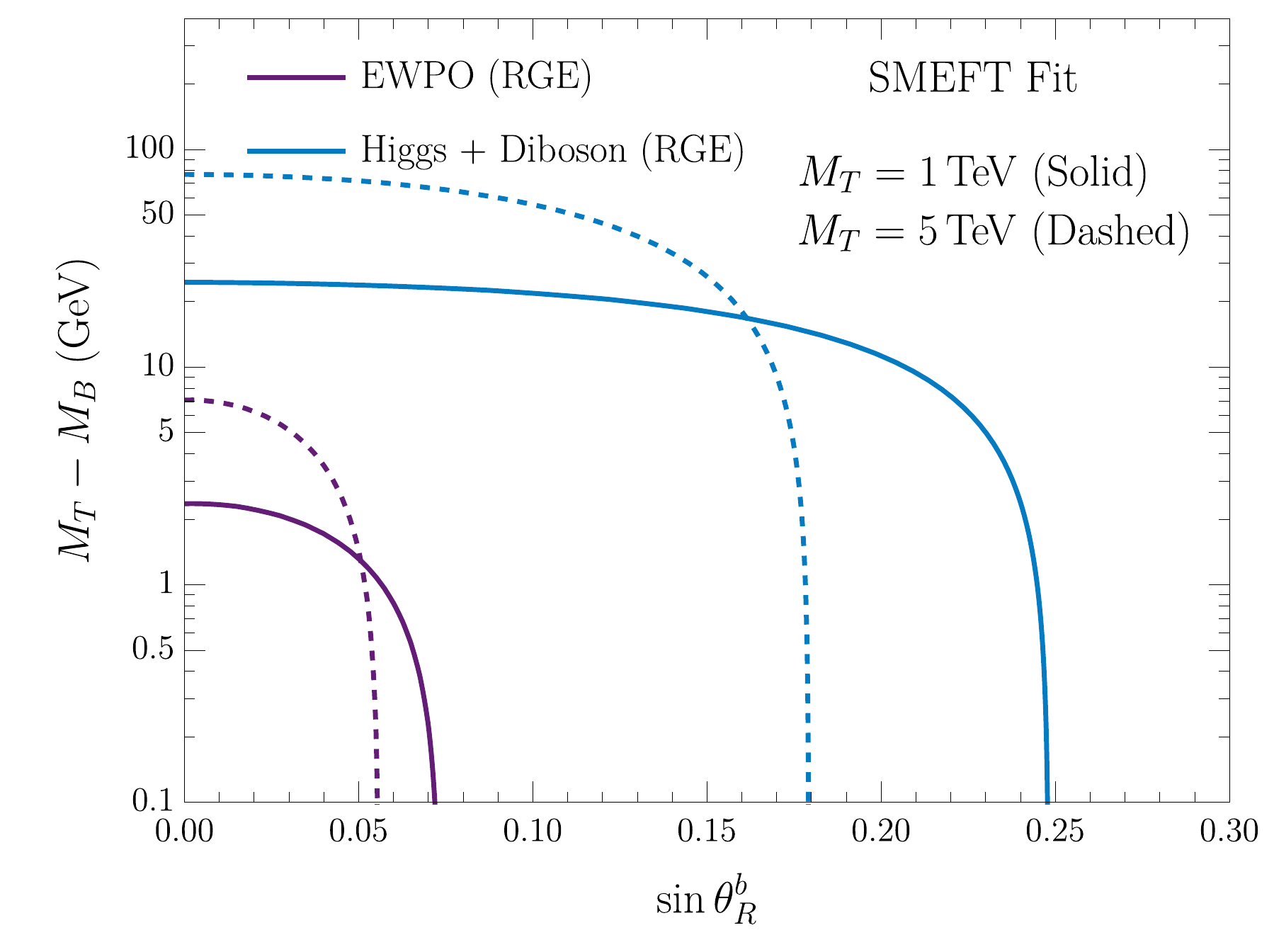}
\vskip -0.25cm
\caption{LHS:  $95\%$ confidence level  limits on $C_{Ht}$ and $C_{Hb}$ when the other coefficients are set to the correlated values of the $(T\,B)$ VLQ model with $M_T=M_B$  at the matching scale and including RG  evolution of the coefficients to $M_Z$.  The magenta
and yellow lines correspond to predictions for the coefficients with $\delta M_{TB}=0$ and $\delta M_{TB}=10~\textrm{GeV}$.
The coefficients on the axes are evaluated at $\Lambda=1\,\textrm{TeV}$.
RHS: Limits on the mass splitting of the $(T\,B)$ VLQ using the model correlations given in Table \ref{tab:sumlag}.
}
\label{fg:tbvlqmodel}
\end{figure}

In Fig.~\ref{fg:tbvlqmodel} LHS, we directly compare the EWPO and Higgs plus  diboson constraints by considering the SMEFT fits in the $C_{Ht}$ vs. $C_{Hb}$ plane. To include the effects of all the operators, we again set $C_{tH} = -Y_t\,C_{Ht}$, and similarly set $C_{bH} = -Y_b\, C_{Hb}$.
We also include $\Op_{HG}$ with $C_{HG} = 0.65\,\big(\alpha_s / 8\pi\big)\,C_{Hb}$, as implied by Eq.~\ref{eq:vlqtb_chg} for $M_T=M_B=1\,\textrm{TeV}$.
We see that, even including all of these correlations, the EWPO constraint still sets a superior bound to the Higgs plus  diboson data.

In contrast with the $T$ VLQ, the $SU(2)_L$ doublet VLQ  model has two independent parameters in the decoupling limit, the two mixing angles $\sin\theta_R^b\equiv s_R^b$ and $\sin\theta_R^t\equiv s_R^t$, or equivalently, $s_R^b$ and the mass splitting, $\delta M_{TB} = M_T - M_B$ (see Appendix~\ref{sec:vlqdoub} for details).
In the limit $\delta M_{TB} = 0$, the mixing angles are identical, and $C_{Hb} = C_{Ht}$. This is indicated by the magenta line in Fig.~\ref{fg:tbvlqmodel} LHS.
Allowing for a nonzero mass splitting, however, shifts this relation, as can be seen by the yellow line in Fig.~\ref{fg:tbvlqmodel} LHS for $\delta M_{TB} = 10\,\textrm{GeV}$.
It is obvious from this shift that the EWPOs set a strong constraint on the mass splitting for fixed $s_R^b$ via the RGE induced $C_{HD}$.
This is simply a manifestation of the strong constraint on custodial symmetry violation, which we comment more on in Appendix~\ref{sec:vlqdoub}.
The behavior in Fig.~\ref{fg:tbvlqmodel} also illustrates that varying the two mixing angles sweeps out a region in the $C_{Ht}$ vs. $C_{Hb}$ plane, but that there is still a very tight relationship between all five operators generated  by the model, which changes the interpretation of the global fits in this context significantly.

On the RHS of Fig.~\ref{fg:tbvlqmodel}, we reinterpret the SMEFT bounds in the $\delta M_{TB}$ vs. $s_R^b$ plane, showing both the EWPO constraint (purple) and the Higgs plus  diboson constraint (blue).
We show the results for both $M_T = 1\,\textrm{TeV}$ (solid) and $M_T = 5\,\textrm{TeV}$ (dashed), and note that the logarithmic dependences on $\delta M_{TB}$ and $s_R^b$ have opposite signs.

\subsection{Global Fit to SMEFT Coefficients}
\label{sec:global}

\begin{figure}[t]
\centering
\includegraphics[width=0.9\linewidth]{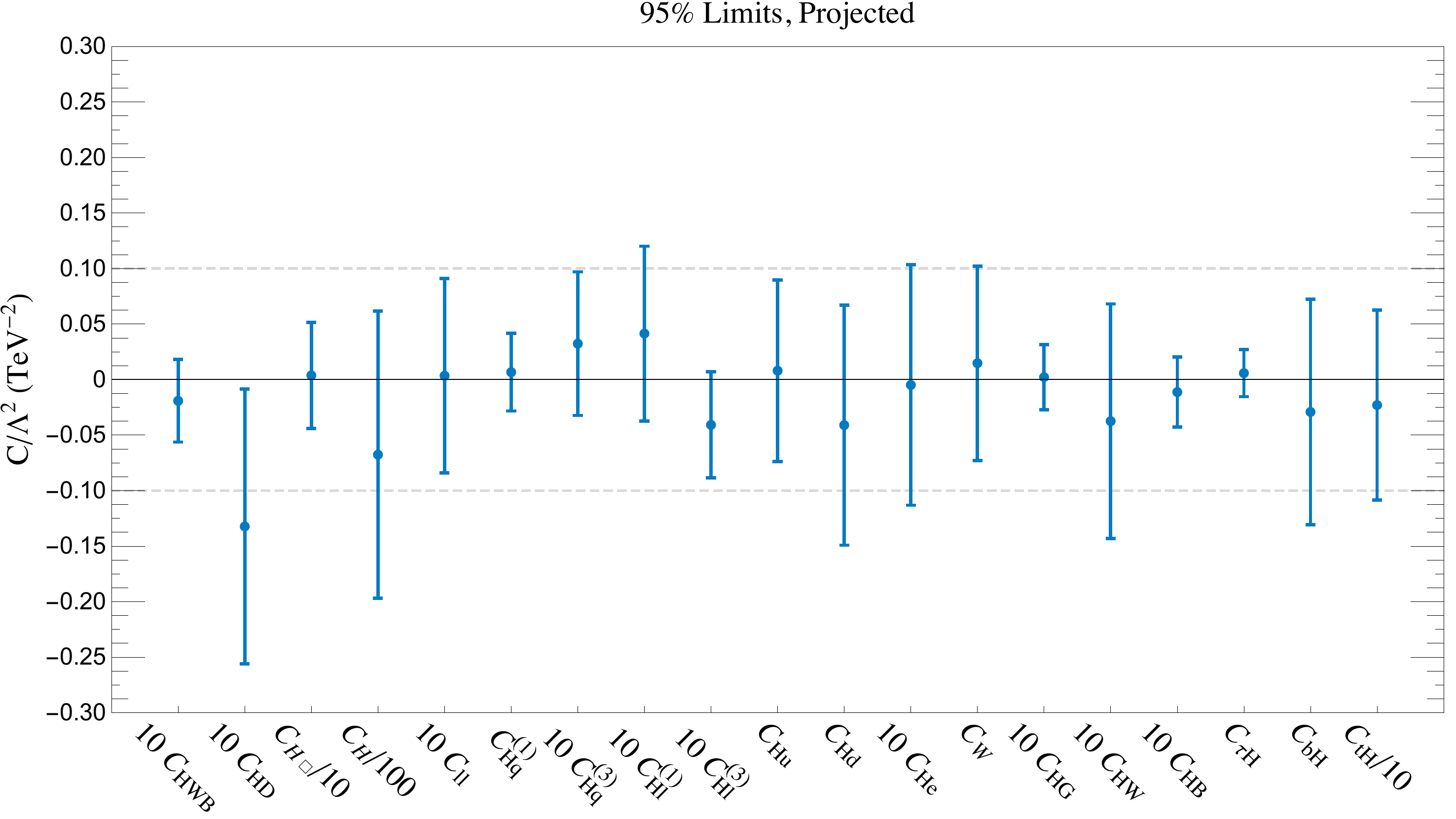}
\caption{95\% C.L. limits on the  coefficients of operators, with all other coefficients set to zero. The bounds on operators involving fermions
assume universal coefficients, except for $C_{bH}$, $C_{tH}$, and $C_{\tau H}$, which modify only the third-generation Yukawa couplings.}
\label{fig:warproj}
\end{figure}

\begin{figure}[hbp!]
\centering
\includegraphics[width=0.9\linewidth]{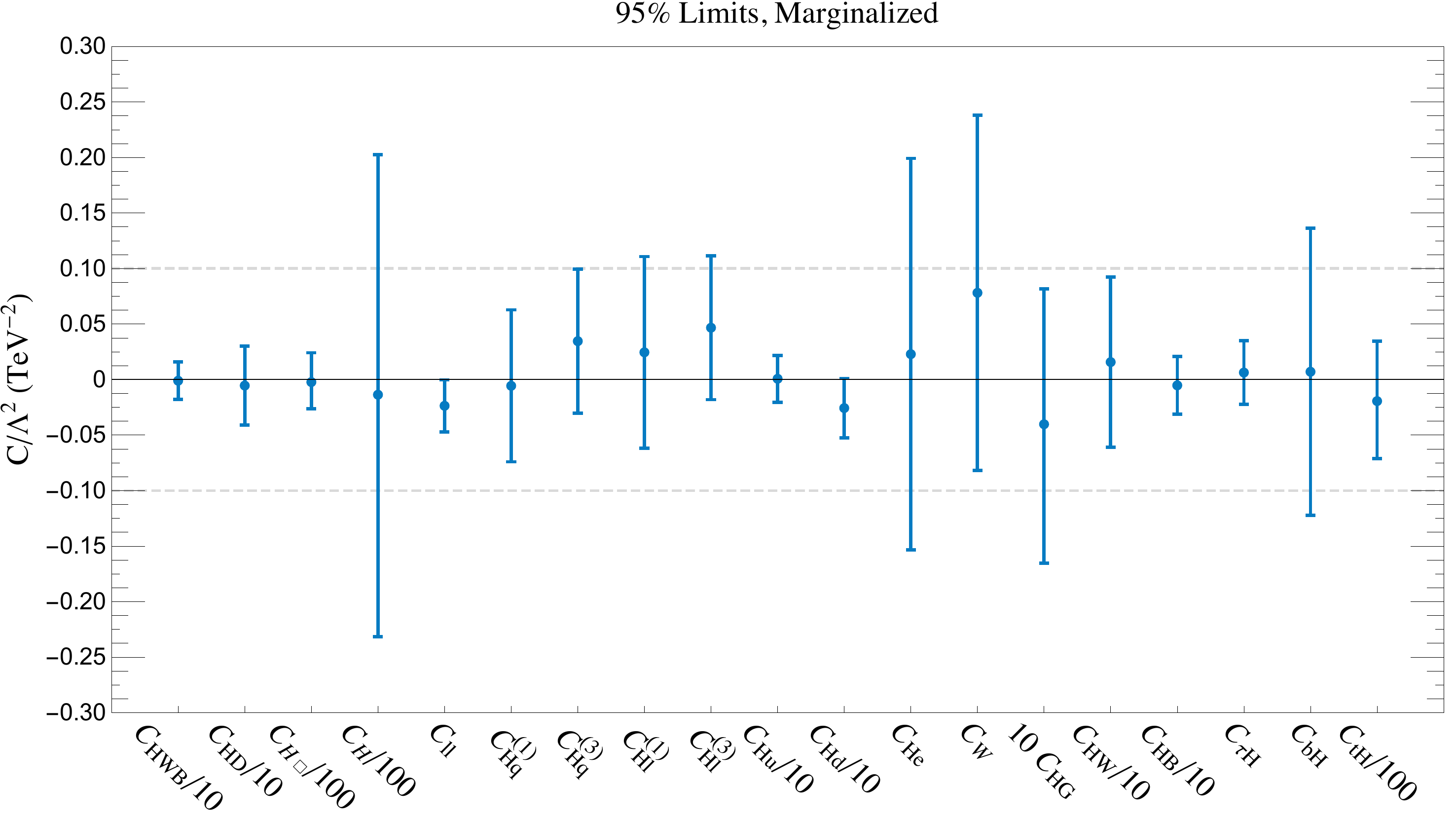}
\caption{
95\% C.L. limits on each the coefficients of each operator, marginalizing over all other operators. We assume universal coefficients for operators involving fermions, except for $C_{bH}$, $C_{tH}$, and $C_{\tau H}$, which modify only the third-generation Yukawa couplings.}
\label{fig:warmarg}
\end{figure}

As a by-product of our study, we present an updated global fit to the 19 SMEFT coefficients considered here in the Warsaw basis.
In comparison to Ref.~\cite{Ellis:2018gqa}, this fit includes higher integrated luminosity data from ATLAS~\cite{Aad:2019mbh} and CMS~\cite{CMS:2020gsy}, as well as the NLO QCD corrections to $VV$ and $VH$ production with the full distributions as in Ref.~\cite{Baglio:2020oqu}.
The results are shown in Fig.~\ref{fig:warproj} with each coefficient treated individually and in Fig.~\ref{fig:warmarg} when marginalizing over all the couplings.  
The result for $C_{Hq}^{(3)}$ is considerably strengthened from that  of Ref.~\cite{Ellis:2018gqa}, due to the inclusion of the $WZ$ data not available in that work. .  
In general, the inclusion of the  higher luminosity Higgs data strengthens the limits  that are not dominated by EWPO data by ${\cal O}(20\%)$.
Note that, in contrast to many of our particular model fits, here we assume universal couplings to the quark operators $C_{Hq}^{(1)}$, $C_{Hq}^{(3)}$, $C_{Hu}$ and $C_{Hd}$.
The assumption that the quark couplings are flavor blind has a major effect on the fits, as the constraints now have a significant contribution from diboson  and Higgstrahlung production from first-generation quarks.
Numerical values for the fits can be found at
\url{https://quark.phy.bnl.gov/Digital_Data_Archive/dawson/smeft_20}.

\section{Discussion}
\label{sec:conc}

A major goal of precision measurements at the LHC is to uncover hints of new physics through patterns of deviations from the SM.  In this work,
we examine how fits to SMEFT coefficients that are predicated on patterns of coefficients generated in different UV complete models give
information about the high scale physics. Of particular  interest to us are the assumptions made when forming inferences about the source of 
new physics from SMEFT fits.  

Only two of our models, the $Z_2$ non-symmetric singlet model and the 2HDM generate a shift in the Higgs tri-linear coupling $C_H$.    In the singlet model, this shift is correlated with a non-zero $C_{H\square}$ term that can be observed in $VV$ and $VH$ production.  In the 2HDM, the non-zero $C_H$ is directly proportional to the $C_{fH}$ interaction and a weak limit on $C_H$ is obtained.   The 2HDM and the $(T\,B)$ VLQ models generate $C_{fH}$ terms that can be directly measured in Higgs production at the LHC.  The $(T\,B)$ model also generates $C_{Hf}$ couplings
that are limited by precision $Z$ measurements.   When the fits are performed using the patterns of coefficients predicted in each model, the
results are quite different from the global fit results and  in all cases, the correlated fits deviate significantly from the single parameter fits. 

An interesting feature of our work is the importance of the RGE on the interpretation of the fits.
This suggests that redoing the study with complete one-loop matching would be of interest.
If a model predicts coefficients at the matching scale that generate operators through RGE that are constrained by EWPOs or diboson data, then these coefficients are strongly constrained.
The inclusion of RGE completely changes the interpretation of the fits in these cases. 

\begin{figure}
\includegraphics[width=0.8\linewidth]{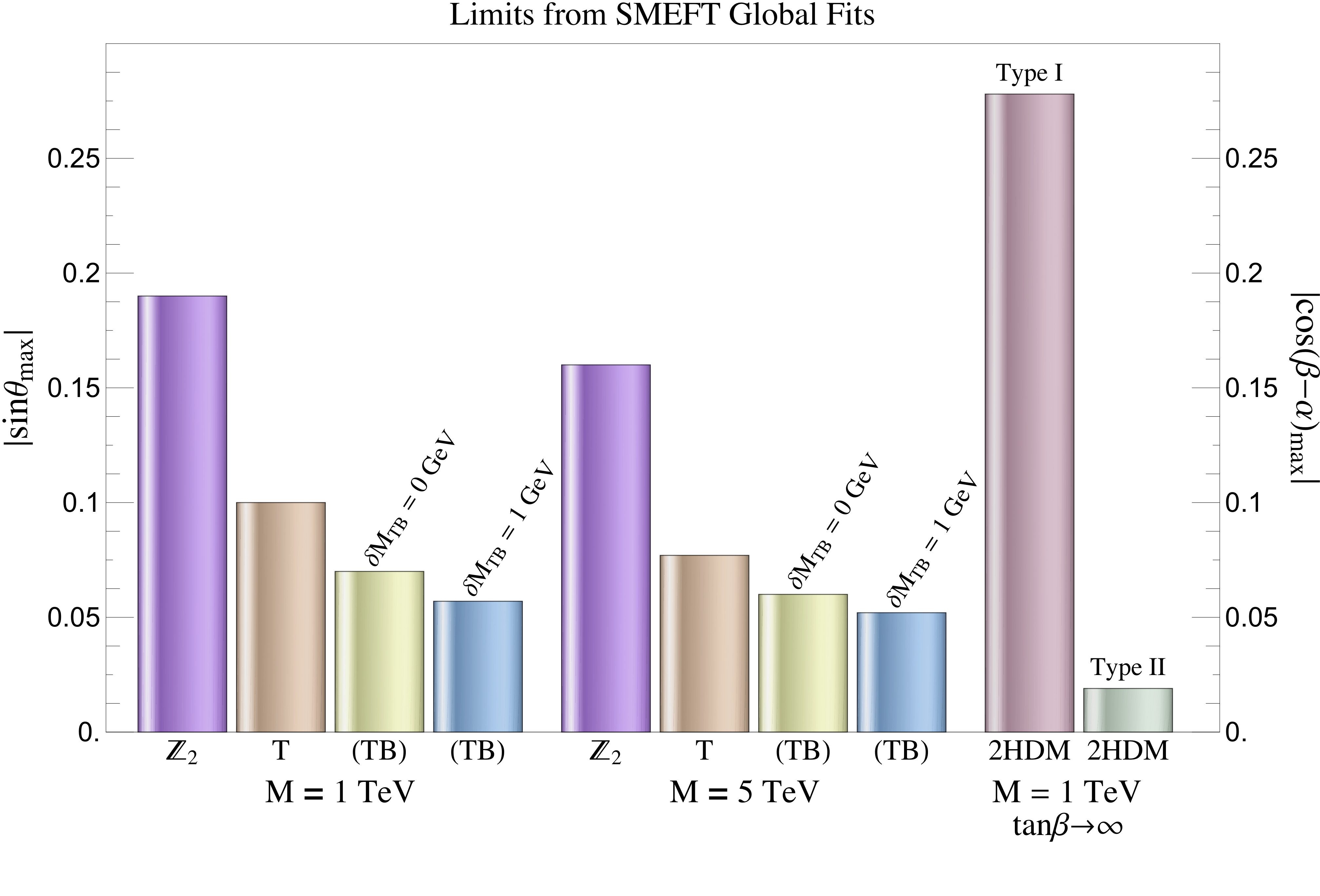}
\vskip -0.8cm
\caption{Maximum allowed mixing angles as a function of the heavy particle mass from the SMEFT global fit. Note that the
2HDM limits are valid in the $\mid\cos(\beta-\alpha)\mid\ll 1$ limit.
}\label{fig:res_sum}
\end{figure}

In Fig.~\ref{fig:res_sum}, we summarize our SMEFT results in terms of the physical parameters of the models and show the maximum
allowed mixing angle from the global fits in each model  as a function of scale.  
We note that these are the limits in the  SMEFT where the heavy particles have been integrated out of the UV complete model.
The fits are sensitive to  the ratios $C_i/\Lambda^2$,  modulo the logarithmic dependence from the RG running. 

Our study is just the beginning of an understanding of the discrimination between UV theories from SMEFT fits~\cite{Barklow:2017suo}.
Follow-up work could include information from top physics~\cite{Durieux:2018ekg,Buckley:2015lku,Cirigliano:2016nyn,Hartland:2019bjb,Durieux:2019rbz,Brivio:2019ius}, a consideration of the importance of the quadratic versus linear SMEFT approximation~\cite{Baglio:2020oqu}, and complete 1-loop matching.  The complete 1-loop matching contributions exist for the singlet 
model~\cite{Ellis:2017jns,Jiang:2018pbd,Haisch:2020ahr}, but not for the VLQ 
models where there are missing heavy-light contributions~\cite{delAguila:2016zcb,Angelescu:2020yzf}. 
The inclusion of flavor observables (such as $B$ and $K$ data) into the fit could further restrict the allowed coefficients in the various 
models~\cite{Alasfar:2020mne,Descotes-Genon:2018foz,Aoude:2020dwv,Aebischer:2020ls,Bobeth:2016llm}. 
It is of considerable interest to expand our study by examining further concrete example models
and performing detailed comparisons of the SMEFT fits with the global fits in the complete models.

\begin{acknowledgments}
SD is supported by the United States Department of Energy under Grant
Contract DE-SC0012704.
The work of SH was supported in part by the National Science
Foundation grant PHY-1915093.
SL is supported by the State of Kansas EPSCoR grant program and the
U.S. Department of Energy, Office of Science, Office of Workforce
Development for Teachers and Scientists, Office of Science Graduate
Student Research (SCGSR) program.
The SCGSR program is administered by the Oak Ridge Institute for
Science and Education (ORISE) for the DOE.
ORISE is managed by ORAU under contract number DE-SC0014664.  Digital data can be found at
\url{https://quark.phy.bnl.gov/Digital_Data_Archive/dawson/smeft_20}.

\end{acknowledgments}

\appendix

\section{Additional Details on the Models} 

\subsection{Singlet Scalar}
\label{sec:scalars}

The most general scalar potential  involving  a real scalar singlet, ${\tilde {S}}$,  and the SM $SU(2)_L$ doublet, $H$, is ,
\begin{eqnarray}
V&=&-\mu_h^2\mid H \mid^2 +{\lambda_h\over 2}(\mid H\mid^2)^2
+{m_S^2\over 2}{\tilde S}^2+ A\mid H \mid^2 {\tilde S}
\nonumber \\ && +{\kappa\over 2}\mid H\mid^2{\tilde S}^2
+{m\over 6}{\tilde S}^3+{\lambda_S\over 24}{\tilde S}^4 .
\label{eq:noz2}
\end{eqnarray}
The parameters can be redefined such that $\langle {\tilde S} \rangle=0$. 
After spontaneous symmetry breaking, the 2 neutral scalars, $H_0=({\tilde h}+v)/\sqrt{2}$ and ${\tilde S}$, mix  to
form the physical scalars, $h$ and $S$,
\begin{eqnarray}
h&=& \cos\theta \,{\tilde h}-\sin\theta \,{\tilde S}\nonumber \\
S&=& \sin\theta \, {\tilde h}+\cos\theta\, {\tilde S}\, ,
\end{eqnarray} 
with  the physical masses, $m_h$ and $M$.  We assume $M \gg m_h$.
The heavy scalar can be integrated out~\cite{Henning:2014wua,deBlas:2014mba,Gorbahn:2015gxa,Dawson:2017vgm}, 
generating  the $2$  operators,
$\Op_H$ and $\Op_{H\square}$ with coefficients,
\begin{eqnarray}
{C_{H\square}\over \Lambda^2}&=&-{A^2\over 2 m_S^4}\nonumber \\
{C_H\over \Lambda^2} &=&{A^2\over 2 m_S^4}\biggl({m A\over 3 m_S^2}-\kappa\biggr)\, . 
\label{eq:z2}
\end{eqnarray}

In terms of the physical parameters of the theory ($m_h, M , \sin\theta$),
\begin{eqnarray}
{C_{H\square}\over \Lambda^2} &=&-{1\over 2v^2 }\biggl({(m_h^2-M^2)\sin (2\theta)\over
m_h^2+M^2 +(M ^2-m_h^2)\cos(2\theta)}\biggr)^2\nonumber\\
&\rightarrow & -{1\over 2v^2}\tan^2\theta\, ,\nonumber \\
C_H &\rightarrow & = -C_{H\square}\biggl(\tan\theta{m\over 3 v}-\kappa\biggr)
\label{eq:singrel}
\end{eqnarray}
where in the last lines, we take the $M \rightarrow \infty$ limit.
Ref.~\cite{Brehmer:2015rna} has pointed out that in some cases,
an improved agreement between the exact (singlet model) UV  theory and the SMEFT can be obtained by retaining the dependence
on $m_h$ in Eq. \ref{eq:singrel}. The Lagrangian parameters $\mu$ and $\kappa$
are limited by the requirement that the minimum of the potential be the electroweak vacuum, (see Fig.
1 of Ref.~\cite{Chen:2014ask}),
\begin{eqnarray}
\mid{m\over v}\mid &<&  6\nonumber \\
\mid \kappa\mid &<& 1\, ,
\label{eq:vacans}
\end{eqnarray}
and so fixing $0<\theta<{\pi\over 2} $,  $\mid C_H\mid \lsim \mid 2\tan\theta+1\mid C_{H\square}$.
In the case where there is a $Z_2$ symmetry, the potential of Eq. \ref{eq:noz2} has $A=\mu=0$.   In this case, there
is a cancellation in Eq. \ref{eq:z2} implying $C_H=0$ and the singlet vev can no longer be fixed to $0$. 

\subsection{A Second Higgs Doublet}
\label{sec:2hdm}

For the 2 Higgs doublet model, we work in the Higgs basis, where the doublets have been rotated such that only the
SM-like doublet, $H_1
$,  gets a VEV, $v$.  In this framework,  the components of the $H_2$ doublet can be taken heavy and we work in the decoupling
limit where the 2HDM can be matched to the SMEFT coefficients~\cite{Gorbahn:2015gxa,Brehmer:2015rna,Belusca-Maito:2016dqe,Corbett:2017ieo}. 
The scalar potential is~\cite{Davidson:2005cw}, 
\begin{eqnarray}
V_S&=& M_1^2\mid H_1\mid^2+M_2^2\mid H_2\mid^2+(Y_3 H_1^\dagger H_2+h.c.)\nonumber \\
&&+{1\over 2} (Z_1\mid H_1\mid^4+Z_2\mid H_2\mid^4)+Z_3\mid H_1\mid^2 \mid H_2\mid^2+Z_4(H_1^\dagger H_2)(H_2^\dagger H_1)
\nonumber \\ &&
-\biggl({Z_5\over 2} (H_1^\dagger H_2)(H_1^\dagger H_2)  +Z_6\mid H_1\mid^2(H_1^\dagger H_2)+
Z_7 \mid H_2\mid^2H_1^\dagger H_2 + h.c. \biggr)\, .
\end{eqnarray}
The Yukawa terms are,
\begin{eqnarray}
V_Y&=& 
  Y_u {\overline q}_L {\tilde {H}} _1 u_R 
+   Y_d{\overline q}_L H_1 d_R 
+   Y_e  {\overline l}_L H_1 e_R 
 \nonumber \\ &&
  +  {\eta_u Y_u\over \tan\beta}  {\overline q}_L {\tilde {H}} _2 u_R 
+   {\eta_d Y_d\over \tan\beta} {\overline q}_L H_2 d_R 
+   {\eta_e Y_e\over \tan\beta} {\overline l}_L H_2 e_R 
+ \text{h.c.} \, ,
\end{eqnarray}
where  $Y_f={\sqrt{2}m_f\over v}$, ${\tilde H}_i=i\sigma_2H_i^*$,   and  the parameters $\eta_f$ depend on the type of 2HDM and are given in Table \ref{tab:2hdm}.  In general, the Yukawa couplings are $3\times 3$ matrices, but we will always take them diagonal when considering
the 2HDM. 
\begin{table}
\centering
\begin{tabular}{|c|c|c|c|}
\hline
 		& \hspace{0.7cm}$\eta_t$\hspace{0.7cm}	
		& \hspace{0.7cm}$\eta_b$\hspace{0.7cm} 
		& $\eta_\tau$\hspace{1.2cm}\\
 \hline\hline
 Type-I & $1$ & $1$ & $1$\\
 \hline
 Type-II  &$1$ &  $-\tan^2\beta $ & ~$-\tan^2\beta$~  \\
 \hline
 ~Lepton-specific~ & 1 & 1 & ~$-\tan^2\beta$~\\
 \hline
 Flipped &$1$  & $-\tan^2\beta$ &  $1$	\\
 \hline
 \end{tabular}
\caption{2HDM couplings of fermions. }
\label{tab:2hdm}
\end{table}

We work in the limit $Y_3/M_2^2 \ll 1$ and
integrate  out the heavy doublet $H_2$ following Refs.~\cite{Henning:2014wua,Belusca-Maito:2016dqe}.  Since the equations of motion imply $Y_3=-Z_6v^2/2$, we
also have $\mid Z_6 v^2/M_2^2\mid \ll 1$.  In the decoupling limit, the heavy masses $M_H\sim M_A\sim M_{H^+}\sim M$  and
the tree level SMEFT contributions are,\footnote{We assume all couplings are real
and neglect flavor indices.}
\begin{align}
&{ C_H\over \Lambda^2}  = 	\frac{Z_6^2}{M^2}  
& {C_{uH}\over \Lambda^2} =   \frac{\eta_u Y_u Z_6 }{\tan\beta M^2} \nonumber \\
& {C_{dH}\over \Lambda^2} =    \frac{\eta_dY_d Z_6 }{\tan\beta M^2} 
& {C_{eH}\over \Lambda^2} =  \frac{\eta_eY_e  Z_6 }{\tan\beta M^2}  \, .
\end{align}
The 2HDM also generates 4-fermi interactions that do not contribute to our tree- level study.
In the decoupling limit,  
\begin{align*}
\cos (\beta - \alpha) \approx
- \frac{Z_6 v^2}{M^2}\, .
\end{align*}
Keeping only third generation fermion masses non-zero,
\begin{align}
& {v^2 C_{H}\over \Lambda^2}   =   \frac{ \cos^2(\beta - \alpha) M^2 }{ v^2}
& {v^2 C_{tH}\over \Lambda^2} =  - \frac{\eta_t Y_t \cos (\beta - \alpha) }{ \tan \beta} \nonumber \\
& {v^2 C_{bH}\over \Lambda^2} = - \frac{\eta_b Y_b \cos (\beta - \alpha) }{ \tan \beta}
& {v^2 C_{\tau H}\over \Lambda^2} =  - \frac{\eta_\tau Y_\tau \cos(\beta-\alpha)}{  \tan\beta}  \, .
\end{align}
Note that we need $\cos(\beta-\alpha){M^2\over v^2}$ to be small for decoupling~\cite{Gunion:2002zf}.

\subsection{Colored Extensions of the SM}
\label{sec:colored}

Finally, we consider extending the Standard Model with new colored fields. In particular, we will consider heavy vector-like quarks, either a singlet or doublet under $SU(2)_L$, and colored triplet scalars. 

\subsubsection{SU(3) Triplet $SU(2)_L$ Singlet Fermion}
\label{sec:vlqsing}
The $T$ VLQ model has  a charge ${2\over 3}$ color triplet, $SU(2)_L$ singlet fermion.  The particles in the  top sector 
 are,
\begin{equation}
\psi_L, T_R^1, T_L^2, T_R^2\, ,
\end{equation}
where $\psi_L, T_R^1$ are the  SM-like left handed quark doublet and right-handed charge ${2\over 3}$ quark and $T^2$ is the new vector-like quark.
The relevant portion of the Lagrangian is,
\begin{eqnarray}
V_Y&=&\lambda_2 {\overline{\psi}}_L{\tilde{H }}T_R^1+\lambda_3{\overline{\psi}}_L{\tilde {H}}T_R^2
+\lambda_5 {\overline {T}}_L^2 T_R^2+h.c.\,
\end{eqnarray}
which can be expressed in terms of   
the physical parameters, $m_t,\, M_T,\, \sin\theta_L^t\equiv s_L^t$.   After the mixing, the physical  fermions are,
\begin{eqnarray}
t_L&=& \cos\theta^t_L \, T_L^1-\sin\theta^t_L \,T_L^2\nonumber \\
T_L&=& \sin\theta^t_L \, T_L^1+\cos\theta^t_L\, T_L^2
\end{eqnarray} 
and we define $q_L^3=(t_L, b_L)^T$ to be the physical  third generation fermion doublet.   (Note that the mixing in the right-handed
quark sector can be rotated away, so there is only one mixing angle in this model.) 

In order to obtain decoupling, the Yukawa interactions must be much smaller than the Dirac mass term,
$\lambda_2 v,~ \lambda_3v \ll \lambda_5$.  In this limit~\cite{Dawson:2012di},
\begin{equation}
s_L^t\rightarrow {v\lambda_3\over\sqrt{2}M_T}\, ,
\label{eq:dect}
\end{equation}
 and 
\begin{eqnarray}
\lambda_2&\rightarrow & {\sqrt{2} m_t\over v}\biggl[1+{(s_L^t)^2\over 2}\biggl({M_T^2\over m_t^2}-1\biggr)\biggr] \sim Y_t\nonumber \\
\lambda_3&\rightarrow & {\sqrt{2} M_T\over v} s_L^t\nonumber\\
\lambda_5&\rightarrow & M_T\biggl[1+{(s_L^t)^2\over 2}\biggr({m_t^2\over M_T^2}-1\biggr) \biggr]\, .
\end{eqnarray}
Hence,  decoupling requires $(s_L^t)^2\sim  {m_t^2\over M_T^2}$ as seen in Eq. \ref{eq:dect}.  

The SMEFT coefficients that are generated at tree level are, 
\begin{eqnarray}
{v^2\over \Lambda^2} \biggl(C_{Hq}^{(1)}\biggr)_{33}&=&{\lambda_3^2v^2\over 4 M_T^2}=-{v^2\over \Lambda^2} \biggl(C_{Hq}^{(3)}\biggr)_{33} 
\nonumber \\
{v^2\over \Lambda^2}C_{tH}&=&{\lambda_2\lambda_3^2v^2\over 2 M_T^2} \, .
\label{eq:topsing}
\end{eqnarray}
It is clear that there is only $1$ independent SMEFT coefficient in this model at tree level.  

The $T$ VLQ model generates a contribution to $\Op_{HD}$ through the running of $C_{Hq}^{(1)}$,
 \begin{equation}
 {\dot{C}}_{HD}=-24 \biggl(C_{Hq}^{(1)}\biggr)_{33} (Y_b^2-Y_t^2)\, .
 \end{equation}
 Neglecting the $b$ mass, matching at $\Lambda=M_T$,  ${(C_{Hq}^{(1)})_{33}(M_T)\over\Lambda^2}={(s_L^t)^2\over 2}$,  and evolving
 to $m_t$,  we find
 \begin{eqnarray}
 \Delta T_{EFT} &=& -{v^2\over 2 \alpha}C_{HD}(m_t)\nonumber\\
 &=& 2 (s_L^t)^2 T_{SM} \log\biggl({M_T^2\over m_t^2}\biggr)\, , \quad T_{SM}={3\over 16 \pi s_W^2}{m_t^2\over M_W^2} \, ,
 \label{eq:teft}
 \end{eqnarray}
 reproducing the logarithmic contribution of the UV complete $T$ VLQ model.  The complete model, however,
 has the $s_L^t\rightarrow 0$, $M_T\rightarrow \infty$ limit~\cite{Dawson:2012di,Lavoura:1992np},
 \begin{equation}
 \Delta T_{UV}=T_{SM} \biggl[-2+(s_L^t)^2 {M_T^2\over m_t^2}+2 \log\biggl({M_T^2\over m_t^2}\biggr)\biggr]
 \label{eq:tuv}
 \end{equation}
 and we note that the SMEFT cannot reproduce the (numerically significant) $(s_L^t)^2 M_T^2/m_t^2$ term of the UV complete model.

The $T$ VLQ generates $\Op_{HG}$ at 1-loop,
\begin{equation}
{v^2\over \Lambda^2} C_{HG}= {\alpha_s\over 8\pi}(s_L^t)^2\biggl(F_{1/2}(M_T)-F_{1/2}(m_t)\biggr)
\end{equation}
where 
\begin{eqnarray}
F_{1/2}(\tau)&=& \tau \biggl[1+(1-\tau)f(\tau)\biggr]\nonumber \\
f(\tau)&=& \biggl[\sin^{-1}\biggl({1\over\sqrt{\tau}}\biggr)\biggr]^2\qquad\tau \gg 1
\label{eq:tridef}
\end{eqnarray}
and $\tau=4m^2/m_h^2$, where $m=m_t, M_T$ and $F_{1/2}\rightarrow {2\over 3}$ in the $m_t\rightarrow \infty$ limit.  
In the high energy limit, the contribution of $C_{HG}$ to Higgs production is highly suppressed by
the cancellation between the top loop and the $T$ loop and there is only a very slight dependence on $M_T$,
\begin{equation}
{v^2\over \Lambda^2} C_{HG}\sim -{\alpha_s\over 8\pi}(s_L^t)^2 {7m_h^2\over  180 m_t^2}\biggl(1-{m_t^2\over M_T^2}\biggr)\, .
\end{equation}

\subsubsection{SU(3) Triplet $SU(2)_L$ Doublet Fermion}
\label{sec:vlqdoub}
We next consider a model with an $SU(2)_L$ doublet and color triplet pair of  vector-like fermions.  We term this
the $(TB) $ VLQ model~\cite{Dawson:2012di,deBlas:2017xtg,Aguilar-Saavedra:2013qpa}. 
The   third generation quarks in the $(T,B)$ model are, 
\begin{eqnarray}
\psi_L^T=(T_L^1,B_L^1), T_R^1, B_R^1, \chi_L^T&=&(T_L^2,B_L^2), \chi_R^T=(T_R^2,B_R^2)\, 
\end{eqnarray}
corresponding to the scalar potential,
\begin{eqnarray}
V_S&=&
\lambda_t {\overline{\psi}}_L{\tilde {H}}T_R^1
+\lambda_b {\overline{\psi}}_L{ {H}}B_R^1
+\lambda_4 {\overline{\chi}}_L{\tilde{H}}T_R^1+\lambda_5 {\overline{\chi}}_L H B_R^1 +M
{\overline{\chi}}_L \chi_R+h.c.
\end{eqnarray}
(A term ${\overline{\psi}}_L\chi_R$ can be rotated away by a redefinition of the fields.)

Diagonalizing the mass matrices requires 4 angles in the left- and right- handed $t-T$  and $b-B$ sectors,
 $\theta_L^t,~\theta_R^t,~\theta_L^b, \theta_R^b$.
Since there are $5$ terms in the Lagrangian, there are $5$ independent parameters which we take to be the
physical masses and one mixing angle,
\begin{equation}
M_T, m_t, M_B, m_b, s_R^b\, .
\end{equation}
For small mixing angles,
\begin{eqnarray}
(s_L^t)^2&\sim& {m_t^2\over M_T^2} (s_R^t)^2 \nonumber \\
(s_L^b)^2& \sim & {m_b^2\over M_B^2} (s_R^b)^2 \, .
\end{eqnarray}
The mixing in the right-handed top sector is determined from that in the right-handed bottom sector,
\begin{equation}
(s_R^t)^2=(s_R^b)^2\biggl({M_B^2-m_b^2\over M_T^2-m_t^2}\biggr)+\biggl({M_T^2-M_B^2\over M_T^2-m_t^2}\biggr)\, .
\label{eq:massdif_tb}
\end{equation}
 If the mass splitting  between  the $T,B$ particles is small, $\delta M_{TB}=M_T-M_B, ~{\mid \delta M_{TB}\mid\over M_T}  \ll 1$, and $m_t \ll M_T$, the mixing
 in the top sector is, 
\begin{equation}
(s_R^t)^2=(s_R^b)^2+(c_R^b)^2{2\delta M_{TB}\over M_T}\, .
\label{eq:massdif}
\end{equation}
For small mixing angles, and $m_t \ll M_T$, $m_b \ll M_B$,~\cite{Cacciapaglia:2010vn}
 \begin{eqnarray}
 s_R^t&=& {\lambda_4 v\over \sqrt{2} M_T}\nonumber \\
 s_R^b&=& {\lambda_5 v\over \sqrt{2} M_B}\, . 
 \end{eqnarray}
 
 At tree level, the $(TB)$ doublet model  generates $\Op_{Ht},~\Op_{Hb},~\Op_{Htb}, \Op_{tH},~\Op_{bH}$~\cite{Chen:2017hak},
 \begin{eqnarray}
 {v^2\over \Lambda^2}C_{Ht}&=&-{\lambda_4^2v^2\over 2M_T^2}=-(s_R^t)^2\nonumber \\
 {v^2\over \Lambda^2} C_{Hb}&=&{\lambda_5^2v^2\over 2M_B^2} =(s_R^b)^2\nonumber \\
{ v^2\over \Lambda^2} C_{tH}&=&-{\sqrt{2}m_t\over v} {v^2\over \Lambda^2} C_{Ht}\nonumber \\
 {v^2\over \Lambda^2}  C_{bH}&=&{\sqrt{2}m_b\over v} {v^2\over \Lambda^2} C_{Hb}\nonumber\\
 {v^2\over \Lambda^2} C_{Htb}&=& 2s_R^t s_R^b \, .
 \end{eqnarray} 
 Only $C_{Ht}$ and $C_{Hb}$ are independent and are related by Eq. \ref{eq:massdif} to the heavy masses.  We can consistently
 take $\Lambda=M_T$ or $\Lambda=M_B$.
 From  the measurement of $Z\rightarrow b {\overline {b}}$, the right-handed coupling to the $b$ is small,   $s_R^b< .115$~\cite{ALEPH:2005ab}, corresponding to 
 ${v^2\over \Lambda^2}  C_{Hb} < .013$, independent of $M_B$.    It is therefore consistent to consider the small $s_R^b$ limit.

 Similarly to the $T$ VLQ model, $\Op_{HD}$ is generated from the running of $C_{Hu}, ~C_{Hd}$ and $C_{Hud}$,
 \begin{equation}
 {\dot{C}}_{HD}=-24\biggl[Y_t^2C_{Ht}-Y_b^2 C_{Hb}+Y_b Y_t C_{Htb}\biggr]\, .
 \end{equation}
 Neglecting the $b$ mass and considering small $s_R^b$ and $\mid\delta M_{TB} \mid/M_T$,
 \begin{eqnarray}
 {v^2\over \Lambda^2}{\dot{C}}_{HD}&=& 24 Y_t^2 (s_R^t)^2\nonumber \\
 &\sim & 24 Y_t^2 \biggl[ (s_R^b)^2 \biggl(1-{2\delta M_{TB}\over M_T}\biggr)+{2\delta M_{TB}\over M_T}\biggr]\, .
 \end{eqnarray}
 Even in the $s_R^b\rightarrow 0$ limit, the running of $\Op_{HD}$  yields a contribution to $\Delta T $ proportional
 to the mass splitting,
 giving
 \begin{equation}
 \Delta T_{EFT}\rightarrow 8 T_{SM} {\delta M_{TB}\over M_T}
 \log\biggl({M_T^2\over m_t^2}\biggr)\, ,
 \end{equation}
 reproducing the logarithm of the UV complete model~\cite{Dawson:2012di,Lavoura:1992np}.   Comparing with Eq. 57 of~\cite{Dawson:2012di}
 we see that  $\Delta T_{EFT}/\Delta T({\text{full}})\sim 1.4$, implying that the limits obtained in the EFT will be more stringent
 than the actual limits in the full theory. 
 
 At one-loop,  $\Op_{HG} $ is generated, 
 \begin{eqnarray}
 {v^2\over \Lambda^2}  C_{HG}&=&{\alpha_s\over 8 \pi} \biggl((s_R^t)^2\biggl[F_{1/2}(M_T)-F_{1/2}(m_t)\biggr]+(s_R^b)^2F_{1/2}(M_B)\biggr)
 \nonumber \\
 &\sim &{\alpha_s\over 8 \pi}(s_R^b)^2 (.65)
 \end{eqnarray}
 where $M_T=M_B=1\,\textrm{TeV}$ in the last equation.  Note that the cancellation between the SM top quark and the $T$ VLQ contribution that was
 observed in the $T$ VLQ model is weakened due to the presence of two heavy VLQs.

\subsubsection{SU(3) Triplet Scalar}
\label{sec:colorscalar}
Finally, we consider a model with a complex color triplet scalar, $s$, with charge $Q={2\over 3}$.
It is interesting to see how the predictions differ from those of the $T$ VLQ described above.  
The relevant interaction terms are, 
\begin{equation}
L=m_0^2 s^{*A}s^A +{\lambda_s\over 2} (s^{*A}s^A)^2+\kappa s^{*A}s^A\mid \phi^\dagger\phi\mid^2\, ,
\end{equation}
  where $A,B,C=1...8$ are color indices.  The mass of the colored scalar is $m_s^2=m_0^2+{\kappa v^2\over 2}$ and  the parameter $\kappa$  measures the amount of the scalar mass due to electroweak symmetry breaking.  
    At tree level, the model generates $4$ -fermion operators~\cite{deBlas:2017xtg},
$\Op_{qq}^{(1)}=-\Op_{qq}^{(3)}$ 
that do not contribute to our study, but 
enter at tree level in Drell-Yan and
di-jet production at the LHC.  At one loop, the colored scalar generates $\Op_{HG}$,
\begin{equation}
{C_{HG}\over \Lambda^2}={\alpha_s\kappa\over 32\pi m_s^2}F_0(\tau_s)\rightarrow -{\alpha_s\kappa\over 96 \pi m_s^2}
\end{equation}
where $F_0(\tau_s)=\tau_s\biggl[1-\tau_s f(\tau_s)\biggr]$, $\tau_s=4m_s^2/m_h^2$, and $f(\tau)$ is defined in Eq. \ref{eq:tridef}.


\clearpage
\bibliographystyle{utphys}
\bibliography{smeft}

\end{document}